\documentstyle [aps,psfig]{revtex}

\begin{document}
\draft
\title{Oscillatory approach to the singularity in vacuum
spacetimes with $T^2$ isometry}
\author{Beverly K. Berger\thanks{E-mail: berger@Oakland.edu}}
\address{Department of Physics, Oakland University, Rochester, MI 48309
USA}
\author{James Isenberg\thanks{Email: jim@newton.uoregon.edu}}
\address{Department of Mathematics, University of Oregon,
Eugene, Oregon 97403 USA}
\author{Marsha Weaver\thanks{E-mail:
weaver@aei-potsdam.mpg.de}}
\address{Max-Planck-Institut f\"ur Gravitationsphysik,
 Am M\"uhlenberg 1, D-14424 Golm, Germany}
\maketitle
\bigskip
\begin{abstract}
We use qualitative arguments combined with numerical
simulations to argue that, in the approach to the singularity in a
vacuum
solution of Einstein's equations with $T^2$ isometry, the evolution at
a generic point in space is an endless succession of Kasner epochs,
punctuated by bounces in which either a curvature term or a twist
term becomes important in the evolution equations for a brief time.
Both
curvature bounces and twist bounces may be understood within the
context of local mixmaster dynamics although the latter have never been
seen before in spatially inhomogeneous cosmological spacetimes.
\end{abstract}
\pacs{04.20.Dw,98.80.Dr}

\section{Introduction}
Thirty years ago, Misner \cite{misner} and Belinskii,
Khalatnikov and Lifshitz (BKL) \cite{bkl2} noted that Bianchi~IX
spatially homogeneous cosmological solutions of Einstein's
equations seem to exhibit a sort of oscillatory behavior in
the approach to the singularity.  This behavior, labeled
``mixmaster'' \cite{misner}, involves an infinite sequence of
periods (or ``epochs'') during which the solution evolves
essentially as a Kasner spacetime \cite{kasner,harvey}, with
each Kasner epoch ended by a ``bounce'' of short duration
which changes the evolution from that of one Kasner to that
of another one.  The sequence of Kasners satisfies a rule,
called the Kasner map, which takes one Kasner in the sequence
to the next.\footnote{However, it is not necessarily expected
that the evolution converges to a single such sequence of
Kasners.  It may be that the evolution always eventually
diverges from any one such sequence and another sequence, which
again follows the Kasner map, becomes a better approximation.} This 
characterization of the Bianchi~IX singularity has
recently been made rigorous~\cite{ringstrom}.
BKL also made the rather surprising claim
\cite{bkl2,bkl1} that in spatially inhomogeneous solutions of
Einstein's equations, timelike observers\footnote{In this paper
we mean by the term ``observer'' a timelike path with constant
spatial coordinates.  We assume that a foliation and threading
have been chosen.  Whether results of the sort discussed here
will be seen by inequivalent sets of observers is not yet
generally known.  However, this does seem to be true at least
in certain cases \cite{jim-vinceU1}.} approaching a Big Bang
or Big Crunch singularity should generally see this oscillatory
behavior, with the Kasner epoch seen by one observer differing
from that seen by other neighboring observers, but the sequence
of Kasners for each observer still satisfying the Kasner map.

Our previous study of the magnetic Gowdy family of spacetimes
\cite{maggow} provided the first firm support for BKL's claim
in a spatially inhomogeneous setting.  In that work, we
numerically evolve spacetimes in the family using the
standard areal (or ``Gowdy'') time foliation, and we find
that generic observers see oscillatory behavior in the
metric evolution.  Moreover, our studies of the magnetic
Gowdy spacetimes indicate that the sequence of Kasners seen
by each observer follow the pattern of succession predicted
by BKL \cite{bkl2,bkl1}.
Further, these studies agree with the qualitative picture
which the Grubi\v{s}i\'{c}-Moncrief method of consistent
potentials (MCP) suggests \cite{vince-grubisic}.

Since this magnetic Gowdy work, numerical and MCP studies of two
other families of cosmological spacetimes have been carried out:
the $T^2$ symmetric vacuum spacetimes and the $U(1)$ symmetric
vacuum spacetimes.  Both studies strongly support the BKL claim
that the approach to the singularity is oscillatory.  The results
for $U(1)$ symmetric solutions have been reported elsewhere
\cite{beverly-vinceU(1)}.  Here, we discuss the behavior near
the singularity for $T^2$ symmetric vacuum spacetimes.

Since $U(1)$ is a subgroup of $T^2 = U(1) \times U(1)$, the
$T^2$ symmetric vacuum spacetimes are a subfamily of the
$U(1)$ symmetric vacuum spacetimes.\footnote{Note that in
studies of the behavior near the singularity in
$U(1)$ symmetric spacetimes, a restrictive assumption is
made.  This restriction is consistent with the full range
of $T^2$ symmetric solutions.}  One may then ask why it is
useful to study the $T^2$ symmetric family directly.  The
reason is that, since the equations for the $T^2$ symmetric
family are considerably simpler (1+1 PDEs rather than 2+1
PDEs) the numerical studies can be done significantly more
accurately.  Hence, the studies are more accurate for the $T^2$
symmetric family, and the behavior of the bounces seen
by the observers can be monitored more
carefully.  The MCP analysis has been carried out in great detail in
this
simpler case.  We report this study here both to present
the detailed picture it gives of the dynamics in these
spacetimes, and also, we hope, as an aid to obtaining
rigorous results about the dynamics.

We define the $T^2$ symmetric family in Sec.~II, noting
the relationship between this family and others, such as the
Gowdy \cite{gowdy} and the Kasner spacetimes.  Also in Sec.~II, we
discuss the areal function and coordinates, recalling
results which justify their use for $T^2$ symmetric
solutions and writing out the field equations.  In
Sec.~III, we set up the MCP treatment of the evolution
equations for the $T^2$ symmetric spacetimes and use
it to argue that oscillatory behavior occurs.  We recall
that in setting up the MCP form of a given set of evolution
equations, one presumes that at each spatial point, the
fields evolve to Kasner epoch values (not necessarily at
the same time for all spatial points); one then substitutes
these Kasner-like values of the fields into the right hand
side of the evolution equations, and attempts to infer how
the various terms in these equations should behave in time,
and what the resulting behavior of the fields should be.
This analysis predicts that there should be three types of
bounces in $T^2$ symmetric spacetimes: curvature bounces,
twist bounces and kinetic bounces.  A kinetic bounce is
not a transition between two distinct Kasner epochs.
Rather, it occurs within a Kasner epoch.  However, in
terms of the evolution of the metric functions, it is a bounce on a par
with the others and its occurrence is necessary for the
oscillatory behavior to continue.  We state in Appendix~A
the explicit evolution of the fields during each of the
three bounces (ignoring in each case terms in the evolution
equations which are small) and discuss the qualitative
nature of each in Sec.~III.  We compare the MCP predictions
for bounce behavior with those of BKL.  We also discuss in
Sec.~III the MCP argument that, in these spacetimes, an
observer following a timelike path of constant spatial
coordinate should see an unending succession of bounces,
a key ingredient of mixmaster dynamics and the BKL claims.

MCP analysis provides useful predictions, but is limited in
that, besides being nonrigorous, it does not predict whether
generic initial data will evolve into a spacetime in which,
along each appropriate timelike observer's path, a Kasner-like
state is reached.  (This, again, is a prerequisite for
carrying out the MCP study.)  To justify the MCP predictions,
we rely on numerical studies of $T^2$ symmetric solutions.
For representative
sets of initial data, Kasner epoch values for the fields are
reached at each spatial point.  Once the Kasner regime is reached at a
given spatial point, the bounces occur as predicted by our MCP studies,
as far as we are able to carry out the evolution.  Discussion 
of these results is presented in Sec.~IV.  
These numerical studies do not prove that $T^2$ symmetric
solutions generically exhibit oscillatory behavior near the
singularity, as predicted by BKL.  They do, however, strongly
support this contention.

In the magnetic Gowdy family of spacetimes, we have found that in a generic
solution, conditions can occur at nongeneric spatial points (e.g., the
derivative of a metric component has a zero) with the result that at various
points near this nongeneric point, there is only a finite number of bounces.
While similar conditions occur at nongeneric spatial points in a generic
$T^2$ symmetric solution, and while these conditions again appear to persist,
there is as yet no evidence that the bounces stop at nearby points, in
contrast to the situation in a magnetic Gowdy solution.
The occurrence of the exceptional points is observed in
the numerical simulations, and the long time behavior is
predicted by the MCP analysis.  We discuss and exhibit
exceptional points in Sections~III and IV, but leave extensive
discussion of these to future work \cite{longmaggow}.
We make concluding remarks in Sec.~V.


\section{The $T^2$ symmetric spacetimes}
We define the $T^2$ symmetric family of spacetimes to
consist of globally hyperbolic solutions of the vacuum
Einstein equations with compact Cauchy surfaces
and with a $T^2$ isometry group acting spatially and
without fixed points.  Generally for spacetimes in
this family, at least one of the ``twist'' functions
\begin{equation}
\label{twistfunctions}
K_{(X)}\,
:=\epsilon_{\mu \nu \rho \lambda}
X^{\mu} Y^{\nu} \nabla^{\rho} X^{\lambda}
\hspace{20pt} \mbox{ and } \hspace{20pt}
K_{(Y)} \,
:=\epsilon_{\mu \nu \rho \lambda}
X^{\mu} Y^{\nu} \nabla^{\rho} Y^{\lambda}
\end{equation}
does not vanish.  (Here $X$ and $Y$ are a pair of
Killing fields which generate the $T^2$ isometry group.)
If in fact both twist functions do vanish, then one
obtains the important subfamily of Gowdy spacetimes.

The Gowdy spacetimes have been extensively studied, and
it is
believed\cite{vince-grubisic,gowdyphem,numstudy,numinv,vince-jim,KR,R3,RW}
that they are all Asymptotically Velocity Term Dominated (AVTD). Roughly
speaking, this means that as each observer in a given spacetime
approaches the singularity, she sees at most a finite
number of bounces, and eventually settles into a final Kasner
epoch\footnote{ Asymptotically velocity term dominated behavior is
defined more carefully in \cite{vince-jim,eardley}.} which generally
varies from point to point.  Since the Gowdy spacetimes are fairly well
understood, and since they are a set of measure zero in the full family
of $T^2$ symmetric spacetimes, we shall henceforth
presume that one or both of the twist functions is nonzero, in which
case
the only topology compatible with
$T^2$ symmetric spacetimes is $T^3 \times {\mathcal R}^1$.

It is very useful in studying the properties of the evolution
in a given family of spacetimes to have available a universal
choice of spacetime foliation which exactly covers the maximal
globally hyperbolic development of every spacetime in that
family.  As proven in \cite{globalfoliations}, the ``areal
foliation'' (with corresponding areal coordinates)
serves this purpose for the $T^2$ symmetric solutions.  We recall that
the areal foliation chooses spacelike hypersurfaces which are
invariant under the $T^2$ action (thereby containing complete
orbits of $T^2$) with each leaf of the foliation consisting of
all orbits of a fixed area.  That is, if we let $R : T^3 \times
{\mathcal R}^1 \rightarrow {\mathcal R}^1$ be the function
which assigns to a given spacetime point the area of the $T^2$
orbit which contains that point, then the areal foliation
chooses for its time function some $t \propto R$.  In
\cite{globalfoliations} (see also \cite{piotr}), it is shown
that for every $T^2$ symmetric solution $(T^3 \times
{\mathcal R}, g)$ of the vacuum Einstein equations, (i) such
a function $t$ is indeed timelike, (ii) for every value of
$t \in (t_0,\infty)$ with $t_0 > 0$ ($t_0$ fixed for each
spacetime) the $t$ leaf is indeed a $T^3$ Cauchy surface and
(iii) the $t$-hypersurfaces, with $t \in (t_0,\infty)$,
collectively cover the maximal globally hyperbolic region
of $(T^3 \times {\mathcal R}, g)$.  Hence, the areal foliation
provides the desired universal choice of time for the $T^2$
symmetric spacetimes.  We note that for the Gowdy spacetimes,
$t$ is the familiar Gowdy time.

If we use $(x,y) \in T^2$ as coordinates labeling points on the
$T^2$ isometry group orbits, and use $\theta \in S^1$ as a
coordinate parametrizing distinct orbits, then ${(\theta,x,y,t)}$
serve as universal coordinates for the $T^2$ symmetric spacetimes,
and we may write the generic metric for this family in the form
\begin{eqnarray}
\label{areal}
g & = & e^{2(\nu - U)}(-\alpha \, dt^2 + d \theta^2) \nonumber \\
& & \; \; + \sigma \, e^{2U}[dx +A \, dy
+(G_1 + A \, G_2) \, d\theta + (M_1 + A \, M_2) \, dt]^2 \nonumber \\
& & \; \; +\sigma \, e^{-2U} t^2 [dy + G_2 \, d\theta + M_2 \, dt]^2
\end{eqnarray}
where $U$, $A$, $\nu$, 
$\alpha$, $G_1$,
$G_2$,
$M_1$ and
$M_2$
are functions of $\theta$ and $t$ (independent of $x$ and $y$),
and $\sigma$ is a positive constant.  This form~(\ref{areal})
for the $T^2$ symmetric metrics is used for the analysis in
\cite{globalfoliations}.  Here, to make it easier to compare
the present study of $T^2$ symmetric spacetimes with previous
similar studies of magnetic Gowdy spacetimes \cite{maggow} and
Gowdy spacetimes
\cite{gowdyphem,numstudy,numinv,hernsthesis}, it is useful
to replace the time function $t$ by $\tau=-\ln t$ (one still has an
areal
type foliation) and the metric functions $U$, $A$, $\nu$ and $\alpha$ by
the following equivalent functions.
\begin{eqnarray}
\label{Pvar}
P & = & 2 \, U + \tau\,, \\
\label{Qvar}
Q & = & A\,, \\
\label{muvar}
\mu & = & - 2 \ln \alpha \,,\\
\label{lambdavar}
\lambda & = & 4 \, \nu - 4 \, U + 2 \ln \alpha - \tau\,.
\end{eqnarray}
In terms of these variables, the metric takes the form
\begin{eqnarray}
\label{metric}
g & = &  -e^{(\lambda - 3 \tau) / 2} d \tau^2 +
e^{(\lambda + \mu + \tau) / 2} d \theta^2 \nonumber \\
& & \; \; + 
\sigma \, e^{P - \tau} [dx + Q \, dy +(G_1 + Q \, G_2) \, d\theta
+ (M_1 + Q \, M_2) \, (-e^{-\tau} d\tau)]^2 \nonumber \\
& & \; \; + \sigma \,
e^{-P-\tau} [dy + G_2 \, d\theta + M_2 \, (-e^{-\tau} d\tau)]^2.
\end{eqnarray}
We note, for purposes of comparison, that
the metric for magnetic Gowdy spacetimes is the same
as~(\ref{metric}) except that
$G_1$,
$G_2$,
$M_1$ and
$M_2$
vanish.
If one relaxes the assumption of the $T^2$ isometry to allow
it to be a local isometry, then other spatial topologies in
addition to $T^3$ are possible in the magnetic case or the
Gowdy case but not in the general $T^2$ symmetric case with nonvanishing
twist.  The topology affects the spatial boundary conditions of the functions
$P$ and
$Q$, but not the qualitative behavior of the evolution toward the
singularity \cite{longmaggow,vince-jim,david}.

The Einstein vacuum field equations for the $T^2$
symmetric spacetimes \cite{globalfoliations} naturally
divide themselves into four sets.  The first set
\begin{equation}
\label{twistconst}
\partial_\theta
K_{(X)} \,
= 0 \hspace{20pt},\hspace{20pt}
\partial_\tau
K_{(X)} \,
= 0 \hspace{20pt},\hspace{20pt}
\partial_\theta
K_{(Y)} \,
= 0 \hspace{20pt},\hspace{20pt}
\partial_\tau 
K_{(Y)} \,
= 0
\end{equation}
simply tells us that the twist functions are constant
in space and time (and hence are labeled the ``twist
constants'').  For any given $T^2$ symmetric spacetime,
we may always replace $X$ and $Y$ by a linear combination
of themselves and thereby cause one or the other twist
functions to vanish (but not both).  Hence, without loss
of generality, we may further presume that only one of
the twist constants is nonzero.  We label it $K$.

The next two sets are the constraint equations and
the evolution equations for the metric functions
$\{P,Q,\mu,\lambda\}$.  They involve the twist
constant $K$, but are independent of
$\{G_1,
G_2,
M_1, 
M_2\}$.  We discuss these
equations below.  The last set of equations govern
$\{G_1,
G_2,
M_1,
M_2\}$.
They take the form
\begin{eqnarray}
\label{mg1eq}
\partial_\tau G_1 & = & 
-e^{-\tau} \partial_\theta M_1 + 
Q \, e^{(\mu + 2\lambda + 4P + 6\tau) / 4} K\,, \\
\label{mg2eq}
\partial_\tau G_2 & = &
-e^{-\tau} \partial_\theta M_2 
-e^{(\mu + 2 \lambda + 4P + 6\tau) / 4} K.
\end{eqnarray}
We see from these equations that, once $\{P,Q,\mu,\lambda\}$
have been determined, one obtains
$\{G_1,
G_2,
M_1,
M_2\}$
by choosing $M_1(\theta,\tau)$
and $M_2(\theta,\tau)$ to be
arbitrary functions of $\theta$ and $\tau$, choosing
$\hat{G}_1(\theta)$ and
$\hat{G}_2(\theta)$ as
arbitrary (initial data) functions on $S^1$ and then
integrating Eqs.~(\ref{mg1eq})~--~(\ref{mg2eq}) over $\tau$
to get $G_1(\theta,\tau)$
and $G_2(\theta,\tau)$.  Thus
$\{G_1,
G_2,
M_1,
M_2\}$ are nondynamical fields.
They are essentially ``shift functions'', which determine
how the coordinates $(x,y)$ evolve in $\tau$
and $\theta$.  If $K$ is nonvanishing,
$\{G_1,
G_2,
M_1,
M_2\}$ cannot all vanish
everywhere in spacetime; the symmetry group does
not act orthogonally transitively \cite{deforthtrans}.

The dynamics of the gravitational field in $T^2$ symmetric
spacetimes lie in $\{P,Q,\mu,\lambda\}$.  To study these
fields we find it useful to work in Hamiltonian form.
Letting $\pi_P$, $\pi_Q$, $\pi_\mu$ and $\pi_\lambda$
denote the momenta conjugate to these four fields, we find that
$\pi_\mu$ may be eliminated, that the functions
$\{P,Q,\mu,\lambda,\pi_P,\pi_Q,\pi_\lambda\}$ must satisfy the
constraint
equations
\begin{eqnarray}
\label{t2mu}
\pi_\lambda - {1 \over 2} e^{\mu \over 4} & = & 0\,,\\
\label{t2mom}
\pi_P \, \partial_\theta P + \pi_Q \, \partial_\theta Q +
\pi_\lambda \, \partial_\theta \lambda & = & 0
\end{eqnarray}
and that the evolution equations for
$\{P,Q,\lambda,\pi_P,\pi_Q,\pi_\lambda\}$ can be obtained
by varying the Hamiltonian density\footnote{Note
that H in~(\ref{ham}) is {\it not} a superHamiltonian, and
is {\it not} required to vanish as a consequence of the
constraints.  It is a Hamiltonian (density) corresponding
to the choice of time foliation made for these spacetimes.}
\begin{equation}
\label{ham}
H = {1\over 4 \pi_\lambda} \biggl [ \pi_P^2 +
e^{-2P} \pi_Q^2 + e^{-2\tau} (\partial_\theta P)^2 +
e^{2(P-\tau)} (\partial_\theta Q)^2 \biggr ]  +
\sigma \, \pi_\lambda  e^{{(\lambda + 2P + 3\tau) / 2}} K^2.
\end{equation}
In particular we have
\begin{eqnarray}
\label{eveq}  
\nonumber
\end{eqnarray}
\vspace*{-\baselineskip}
\vspace*{-\baselineskip}
\vspace*{-\baselineskip}
\begin{eqnarray}
\label{pev}
\partial_\tau P & = & {\pi_P \over 2 \pi_{\lambda}} \,,\\
\label{pipev}
\partial_\tau \pi_P & = & {1 \over 2 \pi_{\lambda}} [ e^{-2P} \pi_Q^2 +
e^{-2 \tau} \partial_{\theta \theta}
P - {e^{-2 \tau} (\partial_{\theta } P) (\partial_{\theta }
\pi_{\lambda})
\over  \pi_{\lambda}} - e^{2(P-\tau)} (\partial_{\theta } Q)^2 ]
\nonumber \\ & & \hspace{2in}
- \sigma \, \pi_\lambda
 e^{{(\lambda + 2 P + 3 \tau) / 2}} K^2 \,,\\
\label{qev}
\partial_\tau Q & = & {e^{-2P} \pi_Q \over 2 \pi_{\lambda}} \,,\\
\label{piqev}
\partial_\tau \pi_Q & = & {e^{2(P-\tau)} \over 2 \pi_{\lambda}}
[ \partial_{\theta \theta} Q - {(\partial_\theta Q)
(\partial_\theta \pi_{\lambda} )
\over \pi_{\lambda}} + 2 (\partial_\theta P) (\partial_\theta  Q) ]
\,,\\
\label{lamev}
\partial_\tau \lambda & = & - { 1 \over (2 \pi_{\lambda})^2} [
\pi_P^2 + e^{-2 P} \pi_Q^2 
 + {e^{-2 \tau} (\partial_\theta P)^2 +
e^{2(P-\tau)} (\partial_\theta Q)^2 ]}
\nonumber \\ & & \hspace{2in}
+  \sigma \, e^{{(\lambda + 2 P + 3 \tau) / 2}} K^2 \,,\\
\label{pilamev}
\partial_\tau \pi_\lambda & = & -{1 \over 2}  \sigma \, \pi_{\lambda}
e^{(\lambda + 2 P + 3 \tau) / 2} K^2\,.
\end{eqnarray}
The evolution for the remaining metric function, $\mu$,
follows from the constraint~(\ref{t2mu})
\begin{equation}
\label{muev}
\partial_\tau \mu = -2 \sigma \, e^{(\lambda + 2 P + 3 \tau) / 2} K^2.
\end{equation}

The constraint equations, the Hamiltonian and the evolution
equations for the fields for magnetic Gowdy spacetimes are very
similar to these; the main difference is that the twist terms
in the Hamiltonian density and in the evolution equations are
replaced by magnetic terms, with the exponential coefficient
for the magnetic terms, $e^{(\lambda + \tau) / 2}$, differing
from that for the twist terms,
$e^{(\lambda + 2 P + 3 \tau) / 2}$.  
This difference leads
to interesting consequences, which we discuss in a future work
\cite{longmaggow}.


We have already noted the relationship between the familiar
Gowdy spacetimes and the $T^2$ symmetric spacetimes discussed
here.  The (locally) spatially homogeneous subfamily of the
Gowdy spacetimes consists of the Kasner spacetimes.
The (locally) spatially homogeneous subfamily of the
generic $T^2$ symmetric spacetimes, with non-vanishing twist,
consists of Kasner spacetimes as well.  This may seem
surprising, since for the standard Kasner Killing vectors,
all the twist functions vanish.  However, one verifies that,
in the locally homogeneous subfamily, $X$ and $Y$ (with
nonvanishing twist) are a linear combination (with constant
coefficients) of the three Kasner Killing vector fields.
The coefficients are constant, but since the norms are
changing in time, the angles between the two sets of Killing
vectors are changing in time.  More specifically, consider
two orthonormal spatial bases: $E_i$, made up of eigenvectors
of the extrinsic curvature (the Kasner directions), and $e_i$,
such that each vector in the frame is proportional to a
(local) Killing vector and such that two of the frame vectors
are tangent to the isometry orbits generated by $X$ and $Y$.
Then the relation between the two frames is a time dependent
rotation.

Another subfamily of the $T^2$ symmetric spacetimes is
worth noting.  If one chooses initial data with
$Q(\theta, \tau_0) = 0$ and $\pi_Q(\theta,\tau_0) = 0$,
then $Q(\theta, \tau) = 0$ and $\pi_Q(\theta,\tau) = 0$
for all points in the spacetime development of this data.
(See equations~(\ref{qev}) and~(\ref{piqev}).)  Hence, one
can consider a subfamily, the ``polarized'' $T^2$ symmetric
spacetimes, with the metric coefficient $Q$---and the
corresponding gravitational degree of freedom---turned
off.  The polarized $T^2$ symmetric spacetimes have been
studied using Fuchsian methods, and one finds \cite{jim-satya}
that there are full-parameter sets of these which are AVTD
rather than oscillatory near the singularity.  Thus although
oscillatory behavior is expected to occur generically in
$T^2$ symmetric spacetimes, it is not expected to occur
in either the Gowdy or the polarized subfamilies.

\section{MCP argument for oscillatory behavior}
The method of consistent potentials (MCP) is a systematic
approximation scheme \cite{bgimw} for predicting the behavior
of cosmological solutions of Einstein's equations in the
neighborhood of their singularities.  It is based on a key
assumption, which in practice must be checked numerically.
The consequence of this assumption is a weighting of the
influence of various terms in the Hamiltonian.  To describe
this, it is useful to split the Hamiltonian
density~(\ref{ham}) as follows:
\begin{equation}
\label{split}
H = H_0 + H_{kin} + H_{small} + H_{curv} + H_{twist}
\end{equation}
\addtocounter{equation}{-1}
where
\begin{eqnarray}
\label{ham0}
H_0 & = & {1\over 4 \pi_\lambda} \pi_P^2\,, \\
\label
{hamkin}
H_{kin} & = & {1\over 4 \pi_\lambda} e^{-2P} \pi_Q^2\,, \\
\label{hamsmall}
H_{small} & = & {1\over 4 \pi_\lambda} e^{-2\tau}
(\partial_\theta P)^2 \,, \\
\label{hamcurv}
H_{curv} & = & {1\over 4 \pi_\lambda}
e^{2(P-\tau)} (\partial_\theta Q)^2\,, \\
\label{hamtwist}
H_{twist} & = & \sigma \, \pi_\lambda
e^{{(\lambda + 2P + 3\tau) / 2}} K^2 \,.\\
\end{eqnarray}
The assumption, for a fixed $T^2$ symmetric vacuum solution
$(T^3 \times {\cal R},g)$ with the singularity at
$\tau \rightarrow \infty$,\footnote{While the long time
existence result \cite{globalfoliations} does not show that
$\tau \rightarrow \infty$, this is expected to be the case
generically in this family.  We need $\tau \rightarrow \infty$
to obtain the prediction of an unending sequence of bounces since each
(local) Kasner epoch has a finite duration in $\tau$.} concerns the
momentary values of the fields for large
$\tau$.

\vspace{10pt} \noindent
{\bf Kasner Epoch Assumption (KEA):} \, {\it For each $\theta \in S^1$,
there exists a time $\tau_{(\theta)}$ such that}
\begin{eqnarray}
\label{kascon}
\nonumber
\end{eqnarray}
\vspace*{-\baselineskip}
\vspace*{-\baselineskip}
\vspace*{-\baselineskip}
\begin{eqnarray}
\label{kascona}
-\tau_{(\theta)} \ll 0, \\
\label{kasconb}
-P(\theta,\tau_{(\theta)}) \ll 0, \\
\label{kasconc}
P(\theta,\tau_{(\theta)}) - \tau_{(\theta)} \ll 0, \\
\label{kascond}
\lambda(\theta,\tau_{(\theta)}) + 2 P(\theta,\tau_{(\theta)})
+ 3 \tau_{(\theta)} \ll 0,
\end{eqnarray}
{\it such that}
\begin{eqnarray}
\label{expdom}
\nonumber
\end{eqnarray}
\vspace*{-\baselineskip}
\vspace*{-\baselineskip}
\vspace*{-\baselineskip}
\begin{eqnarray}
\label{expdoma} 
H_0 & \gg & H_{kin}, \\
\label{expdomb}
H_0 & \gg & H_{small}, \\
\label{expdomc}
H_0 & \gg & H_{curv} ,\\
\label{expdomd}
H_0 & \gg & H_{twist},
\end{eqnarray}
{\it and such that the terms in the evolution equations
due to $H_0$ dominate the terms in the evolution equations
due to $H_{kin}$, $H_{small}$, $H_{curv}$ and $H_{twist}$.}

Explicitly, this assumption says that for each
$\theta$-labeled observer in the spacetime, there is a time
$\tau_{(\theta)}$ such that all of the exponential factors
in the Hamiltonian density at $(\theta, \tau_{(\theta)})$
are very small.  (The same factors also appear in the
evolution equations.)  This follows from the
conditions~(\ref{kascona})--(\ref{kascond}).\footnote{Note that the
assumption does {\it not} say that the
conditions~(\ref{kascona})--(\ref{kascond}) hold for all $\tau >
\tau_{(\theta)}$ at $\theta$.}  In addition, this assumption says that at
this time the fields
$\{P,Q,\lambda,\pi_P,\pi_Q,\pi_\lambda\}$ have developed
in such a way that the exponential factors control the
relative size of the various terms in the Hamiltonian
density and also in the evolution equations.\footnote{
Dominance of the exponential factors is discussed in
\cite{maggow} and assumed in ``Assumption A''.  There we
failed to note that there are exceptional situations in which
exponential dominance does not hold.  We address those cases
briefly in this paper and in more detail in \cite{longmaggow}.}

The intent of the KEA is to say that, in
any of these spacetimes, the evolution proceeds in such a way
that each of the $\theta$ = constant observers will, at
some time $\tau_{(\theta)}$ (generally varying from point to
point) reach a Kasner epoch.  This follows immediately from
the evolution equations under the conditions assumed.  One
might worry that it is too restrictive to require that $H_0$
dominate the other terms in the Hamiltonian density, because it
is possible that $H_0$ vanishes during a Kasner epoch, but
then the prediction of the MCP analysis is that $H_0$ will
dominate the other terms in the next Kasner epoch occurring
at that value of $\theta$, so the KEA
will thus be satisfied in this next Kasner epoch.

A consequence of the KEA, combined with the
MCP analysis, is that generally (the exceptions are briefly
discussed later in this Section) the fields evolve in such a
way that they do not counteract any explicit exponential decay
or growth of $H_{kin}$, $H_{curv}$ or $H_{twist}$, or that
of terms in the evolution equations~(\ref{eveq}) derived from
$H_{kin}$, $H_{curv}$ or $H_{twist}$.  That is, besides dominating
the relative size of the various terms in the Hamiltonian
density~(\ref{ham}) and in the evolution
equations~(\ref{pev})--(\ref{pilamev}), the exponential terms dominate the
{\it changes} in relative size of the terms.  So, for example, if
$\lambda(\theta,
\tau_{(\theta)}) + 2 P(\theta, \tau_{(\theta)}) +
3 \tau_{(\theta)} \ll 0$ and ${d \over d \tau} (\lambda(\theta,
\tau_{(\theta)}) + 2 P(\theta, \tau_{(\theta)}) +
3 \tau_{(\theta)})|_{(\theta, \tau_{(\theta)})}> 0$, then it
follows from the KEA that $H_{twist}$ is
small relative to $H_0$ at $(\theta, \tau_{(\theta)})$,
and it follows from this assumption combined with the
evolution equations that $H_{twist}$ is growing in $\tau$.

Numerical results, which we discuss in Sec.~IV, indicate that
the KEA holds for $T^2$ symmetric solutions.

We now consider what happens to the gravitational fields along
a fixed $\theta$ observer path after a Kasner epoch has begun
at some time $\tau_{(\theta)}$.  We presume a fixed spacetime,
and we assume the KEA discussed above.
Examining the evolution equations~(\ref{pev})--(\ref{pilamev}), we find
that the right hand sides of all but~(\ref{pev}) and~(\ref{lamev})
are extremely small.  Hence the variables
$\{Q,\mu,\pi_P,\pi_Q,\pi_\lambda\}$ are essentially constant.
The variables $P$ and $\lambda$ are not constant; but if we set
\begin{equation}
\label{wdef}
w := {\pi_P \over 2 \pi_\lambda}
\end{equation}
we have
\begin{equation}
\label{pkas}
\partial_\tau P = w
\end{equation}
and
\begin{equation}
\label{lamkas}
\partial_\tau \lambda = -w^2 + {\cal O}
\end{equation}
where $\cal O$ indicates terms which, as a consequence of the
KEA, can be neglected.  The function $w$
is essentially constant, so $P$ and $\lambda$ evolve linearly
with $\tau$.  Since all four Hamiltonian potentials,
$\{H_{kin}, H_{small}, H_{curv}, H_{twist}\}$ are negligible,
we call the evolution ``velocity dominated'' when a) the KEA holds, b)
$\{Q,\mu,\pi_P,\pi_Q,\pi_\lambda\}$
are essentially constant (note that this implies that $w$ is
essentially constant), and c) $\partial_\tau \lambda \approx -
(\partial_\tau P)^2 = - w^2$.  To reiterate, part a) implies
parts b) and c), that is, the KEA implies
that the evolution is velocity dominated at $\tau_{(\theta)}$.

This predicted pattern of evolution for the variables
$\{P,Q,\lambda,\mu, \pi_P,\pi_Q,\pi_\lambda\}$  and their
spatial derivatives, none increasing or decreasing faster
than linearly, is consistent with the KEA.  The conditions in the
assumption continue
to hold so long as the Hamiltonian potentials stay small
relative to $H_0$.\footnote{Hence the name for this
analysis: the ``Method of Consistent Potentials.''}  To
see whether or not these potentials do indeed stay
small for $\tau > \tau_{(\theta)}$ at $\theta$, we need
to examine the time derivatives of the exponential
quantities in the expressions~(\ref{split}) for $H_{kin}$,
$H_{small}$, $H_{curv}$  and $H_{twist}$.  We have
\begin{eqnarray}
\label{expder}
\nonumber
\end{eqnarray}
\vspace*{-\baselineskip}
\vspace*{-\baselineskip}
\vspace*{-\baselineskip}
\begin{eqnarray}
\label{expdera}
\partial_\tau (- 2 P ) & = & -2 w, \\
\label{expderb}
\partial_\tau (- 2 \tau ) & = & -2, \\
\label{expderc}
\partial_\tau (2 P - 2 \tau ) & = & 2 ( w - 1 ), \\
\label{expderd}
\partial_\tau \left[ {1 \over 2}\left({\lambda + 2 P  +
3 \tau} \right) \right] & = & -{1 \over 2} w^2 + w + {3 \over 2},
\end{eqnarray}
with $w$ approximately constant in $\tau$.  Clearly the value
of the quantity $w$ is crucial in determining whether each
of the potentials grows or not.  In particular, we have, at
fixed $(\theta,\tau_{(\theta)})$,
\begin{eqnarray}
\label{growthanddecay}
\nonumber
\end{eqnarray}
\vspace*{-\baselineskip}
\vspace*{-\baselineskip}
\vspace*{-\baselineskip}
\begin{eqnarray}
\label{grosthanddecaya}
& & H_{kin} \; \; \mbox{grows if} \; \; w < 0 \; \; 
(\mbox{and} \; \; \pi_Q \neq 0) ,\\
\label{grosthanddecayb}
& & H_{small} \; \; \mbox{decays} \; \; (\mbox{if} \; \;
\partial_\theta P \neq 0), \\
\label{grosthanddecayc}
& & H_{curv} \; \; \mbox{grows if} \; \; w > 0 \; \; (\mbox{and} \; \;
\partial_\theta Q \neq 0) ,\\
\label{grosthanddecayd}
& & H_{twist} \; \; \mbox{grows if} \; \; -1 < w < 3.
\end{eqnarray}
Stating this another way, we have the results listed in
Table~\ref{tpotential}. Note that the conditions within the parentheses
in equations (\ref{grosthanddecaya})--(\ref{grosthanddecayc}) ensure
``generic'' behavior.

It does {\it not} follow from Table~\ref{tpotential} that if the
KEA holds there must be at least one growing
potential at $(\theta,\tau_{(\theta)})$.  To argue that, we need
to assume that both $\pi_Q$ and $\partial_\theta Q$ are nonzero
at $(\theta,\tau_{(\theta)})$.  Since generally these smooth
functions, $\pi_Q$ and $\partial_\theta Q$, are nonzero at a
given $(\theta, \tau)$, we define $(\theta,\tau_{(\theta)})$ to
be {\it generic} if neither vanishes, and {\it exceptional} if
one of them does vanish. Exceptional behavior (of a
different type, to be discussed later) also occurs if $w =
1$. Our
subsequent discussion presumes that $(\theta,\tau_{(\theta)})$
is generic unless explicitly stated otherwise; we discuss the
exceptional cases briefly below and in more detail in
\cite{longmaggow}.

Assuming genericity, it does follow from Table~\ref{tpotential}
that, during velocity dominated evolution, at least one of the
potentials is growing.  Indeed, the growth is exponential in
$\tau$.  To see what affect this has, we need to consider how
the fields evolve with one or more of the potentials $H_{kin}$,
$H_{curv}$ and $H_{twist}$ turned on and hence added to $H_0$.

In discussing what happens when some of the potentials become
significant, it is useful to keep track of the {\it generalized
Kasner exponents}.  These are defined to be the
eigenvalues of the extrinsic curvature, divided by the mean
curvature. It
follows from the definition that the sum of the three
generalized Kasner exponents equals $1$. Defining the quantity
\cite{vince-grubisic,gowdyphem,numstudy,numinv}
\begin{equation}
\label{vdef}
v := {1 \over 2 \pi_\lambda} (\pi_P^2 +
e^{-2 P} \pi_Q^2)^{1 \over 2},
\end{equation}
the generalized Kasner exponents for a $T^2$ symmetric solution are
\begin{equation}
\label{t2kasexp}
\kappa_1= {v^2 -1 \over v^2 + 3} + {\cal O} \; \; \; \;, \; \; \; \;
\kappa_2= {2 - 2 v \over v^2 + 3} + {\cal O} \; \; \; \; ,\; \; \; \;
\kappa_3= {2 + 2 v \over v^2 + 3} + {\cal O}.
\end{equation}
As before, $\cal O$ indicates terms which can be neglected at
$(\theta,\tau_{(\theta)})$ when the KEA is
satisfied, as shown in Appendix B.  Note that the KEA implies $v \approx
|\pi_P| / 2 \pi_\lambda = |w|$,
and therefore that the generalized Kasner exponents are
essentially constant in time.   And furthermore, considering the
expressions~(\ref{t2kasexp}), we see that, in addition to
$\kappa_1 + \kappa_2 + \kappa_3 = 1$ which always is
satisfied, when the KEA holds it is also the case that
$\kappa_1^2 + \kappa_2^2 + \kappa_3^2 \approx 1$.   These are
the necessary and sufficient conditions that a set of three
numbers  be a set of Kasner exponents.  Thus the generalized
Kasner exponents are approximately a set of Kasner exponents
when the KEA holds, and so the KEA does indeed imply that the evolution 
is essentially Kasner at
$(\theta,\tau_{(\theta)})$. 

We recall that the BKL parameter ``$u$'' summarizes the information
in a set of Kasner exponents \cite{bkl2,bkl1}.  Except for the
case $\{0, 0, 1\}$, there always exists a $u \geq 1$ such that
\begin{equation}
\label{kasexpinu}
\kappa_{min}= {-u \over 1 + u + u^2} \; \; \; \;, \; \; \; \;
\kappa_{mid}= {1 + u \over 1 + u + u^2} \; \; \; \; ,\; \; \; \;
\kappa_{max}= {u(1 + u) \over 1 + u + u^2}
\end{equation}
where $\kappa_{min}$ is the smallest of a set of Kasner exponents
$\{\kappa_{min}, \kappa_{mid}, \kappa_{max}\}$, $\kappa_{max}$ is
the largest and $\kappa_{mid}$ is the middle value.  For the
expressions~(\ref{t2kasexp}), the relative magnitude depends on
the value of $v$: for $v < 1$, one has
$\kappa_1 < \kappa_2 < \kappa_3$; for $1 < v < 3$,
one has $\kappa_2 < \kappa_1 < \kappa_3$; and for $3 < v$ one has
$\kappa_2 < \kappa_3 < \kappa_1$.  For these different ranges of
values of $v$, we obtain different expressions for $u$:  for example,
if $3 \leq v$, one has $u = (v-1)/2$.  The expressions for $u$
for each range of values of $v$ are given in Table~\ref{uval}.

To understand what happens to the fields when, as a Kasner
epoch progresses, one or more of the potentials becomes
significant, one can look at the evolution of the fields
for $H = H_0 + H_{kin}$, $H = H_0 + H_{kin} + H_{twist}$,
etc.  According to the MCP results summarized in
Table~\ref{tpotential}, there are five such Hamiltonian
densities which in principle must be considered:
$H = H_0 + H_{kin}$, $H = H_0 + H_{curv}$,
$H = H_0 + H_{twist}$, $H = H_0 + H_{kin} + H_{twist}$ and
$H = H_0 + H_{curv} + H_{twist}$.  In fact, for both $H_{kin}$ and
$H_{twist}$, or both $H_{curv}$ and $H_{twist}$, to become significant
simultaneously, some fine tuning is needed. For example, if the KEA holds
at $(\theta_0, \tau_{(\theta_0)})$ with $1 < w < 3$, then we have (for
constants $A \gg 0$ and $B \gg 0$)
$H_{curv} / H_0 \sim e^{-A +
(2 w - 2) (\tau - \tau_{(\theta_0)})}$
and $H_{twist} / H_0 \sim e^{- B +
(-w^2 + 2 w + 3) (\tau - \tau_{(\theta_0)})/2}$, so they are both growing.
Then, for the given
$A$, $B$,
$\tau_{(\theta_0)}$, and the time $\tau'$ of the next bounce, there is some
value of
$w$ such that the two potentials are equal at $\tau'$. A similar argument may
be made for $-1 < w < 0$ and $H_{kin}/H_0$ and $H_{twist}/H_0$.
We find (see Sec.~IV) in our numerical simulations that simultaneous growth
and action of two potentials, though rare, does indeed occur.

\subsection{Derivation of the Bounce Rules}
If a $T^2$ symmetric spacetime satisfies the KEA with a particular value
of $w$, and if it approaches one of the five types of bounces ($H_0 +
H_{kin}$, $H_0 + H_{curve}$, $H_0 + H_{twist}$, $H_0 +
H_{kin}+ H_{twist}$, $H_0 +
H_{curv}+ H_{twist}$), one would like to determine what the MCP predicts
for the value of $w$ after the bounce is over. One way to do this is to
follow the evolution of the fields through each type of bounce (with
appropriate Hamiltonian) into the post-bounce Kasner epoch, and calculate
the change in $w$ directly. The explicit bounce solutions given in
Appendix A facilitate this approach. Another approach, which we use here,
is based on energy and momentum conservation during the bounces. That is,
using the relevant Hamiltonian density for each type of bounce, and
noting that its conservation requires certain quantitites to be of the
same magnitude---but opposite sign---after the bounce as compared to
before, we can determine how $w$ changes. Note that in carrying out this
approach, it is convenient to consider $\tau(t)$ to be a dynamical
variable, dependent on a new time coordinate $t$. It follows that
$\pi_\tau = -H$, and we can work in terms of a Hamiltonian constraint
density ${\cal C}_0$ rather than $H$, and treat the former as a
superhamiltonian (for unit lapse). For example, for $H = H_0 + H_{kin}$,
the Hamiltonian for a kinetic bounce, we have 
\begin{equation}
\label{c0kin}
{\cal C}_0 = 0 = 2 \pi_\tau\,\pi_\lambda \,+\, {1 \over 2} \left(
\pi_P^2
\,+\,\pi_Q^2\,e^{-2P} \right).
\end{equation}

For a kinetic bounce, $\pi_\tau$ and $\pi_\lambda$ are constants of the
motion while $\pi_P$ changes sign. If we then form the quantity
\begin{equation}
\label{wdef}
w = {{\pi_P} \over {2 \pi_\lambda}},
\end{equation}
we immediately obtain the bounce law for kinetic bounces (where unprimed
quantities are evaluated in the Kasner epoch before the bounce and primed
quantities after the bounce)
\begin{equation}
\label{kinbounce}
w' = -\,w.
\end{equation}

We next consider the dynamics for a curvature bounce determined by 
\begin{equation}
\label{c0curv}
{\cal C}_0 = 0 = 2 \pi_\tau\,\pi_\lambda \,+\, {1 \over 2} \left[
\pi_P^2
\,+\,(\partial_\theta Q)^2\,e^{2(P-\tau)} \right].
\end{equation}
Again, $\pi_\lambda$ is a constant of the motion.
If we define
\begin{equation}
\label{zdef}
z = P - \tau\,,
\end{equation}
then
\begin{equation}
\label{zdot}
\partial_t\,z = \pi_P\,-2\pi_\lambda
\end{equation}
(since the variation of ${\cal C}_0$ yields $\partial_t\,\tau =
2\pi_\lambda$). Now
$\partial_t\,z$ must change sign during the bounce so that
\begin{equation}
\label{prebouncecurv}
(\pi_P \,-\,2 \pi_\lambda)' = -(\pi_P \,-\,2
\pi_\lambda).
\end{equation}
Dividing both sides by the constant $2\pi_\lambda$ and solving for
$w'$ yields the bounce law
\begin{equation}
\label{curvbounce}
w' = 2\,-\,w.
\end{equation}

We now consider the twist bounce governed by $H=H_0+H_{twist}$.  In this
case, we have
\begin{equation}
\label{c0twist}
{\cal C}_0=0 = 2\pi_\tau\,\pi_\lambda \,+\, {1 \over 2} \pi_P^2 +
2\sigma\,\pi_\lambda^2\,
\kappa^2\,e^{(\lambda + 2P + 3\tau)/2}.
\end{equation}
If we evaluate the time derivative of the argument of the exponential in
the twist potential
using the equations of motion obtained from the variation of
(\ref{c0twist}), we find
that
\begin{equation}
\label{twistdot}
\partial_t({{\lambda} \over {2}} + P + {{3 \tau} \over 2}) =
3\pi_\lambda + {\pi_P}
\,-\,{{\pi_P^2} \over
{4\pi_\lambda}}\,+\,\sigma\,\pi_\lambda\,\kappa^2\,e^{(\lambda + 2P +
3\tau)/2}.
\end{equation}
Asymptotically (i.e. when the twist potential may be neglected),
$3\pi_\lambda + {\pi_P}
\,-\,{\pi_P^2} /
(4\pi_\lambda)$ is the momentum associated with the time derivative
(i.e. growth rate)
on the left hand side of (\ref{twistdot}). Thus, it must change sign
during the bounce:
\begin{equation}
\label{twist1}
(3\pi_\lambda + {\pi_P}\,-\,{{\pi_P^2} \over{4\pi_\lambda}})' =
-(3\pi_\lambda +
{\pi_P}\,-\,{{\pi_P^2} \over{4\pi_\lambda}}).
\end{equation}
To find a rule for $w$, we must also recognize (from the equations
of motion) that
$\pi_P \,-\,2\pi_\lambda$ is conserved in a twist bounce, so that
\begin{equation}
\label{twist2}
(\pi_P \,-\,2\pi_\lambda)' = (\pi_P \,-\,2\pi_\lambda).
\end{equation}
If we write both sides of Eqs.~(\ref{twist1}) and (\ref{twist2}) so that
a factor of
$\pi_\lambda$ and $2\pi_\lambda$ respectively is shown explicitly,
divide
Eq.~(\ref{twist1}) by Eq.~(\ref{twist2}) to cancel the $\pi_\lambda$'s
on each side (not
the same of course since the left hand side is after and the right hand
side before the bounce), and identify
$w$, we find that
\begin{equation}
\label{twist3}
\left({{3\,+\,2w\,-w^2} \over {w\,-\,1}} \right)' =
-\,\left({{3\,+\,2w\,-w^2}
\over {w\,-\,1}} \right).
\end{equation}
Eq.~(\ref{twist3}) has the solution (there are two solutions but 
one is trivial)
\begin{equation}
\label{twistbounce}
w' = \left({{w\,+\,3} \over {w \,-\,1}}\right).
\end{equation}
Note that $1 < w \le 3$ maps into $w' \ge 3$ while $-1 \le w < 1$ maps
into $w' \le -1$. Thus the former will always yield a curvature bounce
after the twist bounce while the latter will yield a kinetic bounce
after
the twist bounce.

Before deriving the bounce rules for the combined bounces---those with
either $H = H_0+H_{kin} + H_{twist}$ or $H = H_0+H_{curv} + H_{twist}$---
we wish to make two observations. We first note that the quantity $\pi_P
- 2\pi_\lambda$ picks out a direction in local minisuperspace which is
orthogonal to the twist, so its evolution is essentially unaffected by
the presence or absence of the twist potential. Secondly, we note that
the twist bounce rule (\ref{twistbounce}) may be obtained in a way
different from that used above. Specifically, we find that the evolution
generated by $H = H_0 + H_{twist}$ conserves the ``energy''
\begin{equation}
\label{twistE}
E = \left({{\pi_P^2} \over {4 \pi_\lambda}} \,+ \,\sigma\,\pi_\lambda\,
\kappa^2\,e^{(\lambda + 2P + 3\tau)/2}\, +\, 3\pi_\lambda
\right)' =
\left({{\pi_P^2} \over {4 \pi_\lambda}} \,+ \,\sigma\,\pi_\lambda\,
\kappa^2\,e^{(\lambda + 2P + 3\tau)/2}\, +\, 3\pi_\lambda
\right).
\end{equation}
Then, if we factor out $\pi_\lambda$ appropriately from
Eq.~(\ref{twistE}), and if we divide on the left and right hand sides by
the left and right hand sides of Eq.~(\ref{twist2}), we derive
\begin{equation}
\label{twist4}
\left({{w^2\,+\,3} \over {w \,-\,1}} \right)' =
\left({{w^2\,+\,3} \over {w
\,-\,1}}
\right)\,.
\end{equation}
The nontrivial solution of Eq.~(\ref{twist4}) yields
Eq.~(\ref{twistbounce}).

We now consider the curvature-twist bounce. Analogously to
Eq.~(\ref{twistE}), the evolution generated by $H = H_0 +
H_{curv}+H_{twist}$ conserves the energy quantity
\begin{equation}
\label{twistcurvE}
E ={{\pi_P^2} \over {4 \pi_\lambda}}
\,+\,{{(\partial_\theta Q)^2\,e^{2(P-\tau)}}
\over {4
\pi_\lambda}} \,+\,
\sigma\,\pi_\lambda \, \kappa^2\,e^{(\lambda + 2P +
3\tau)/2}\,-\,\pi_P\, +\,
5\pi_\lambda .
\end{equation}
The quantity $\pi_P\,-\,2\pi_\lambda$ is now not conserved. However, as
noted above, the behavior of $\pi_P - \pi_\lambda$ during a
curvature-twist bounce should match its behavior during a curvature
bounce; so we have Eq.~(\ref{prebouncecurv}) (as can be
verified by considering conserved and monotonic quantities). If we now combine
Eqs.~(\ref{prebouncecurv}) and (\ref{twistcurvE}) as we have described
above for Eqs.~(\ref{twist2}) and (\ref{twistE}), we derive 
\begin{equation}
\label{twistcurv1}
\left( {{w^2 - 2w + 5} \over {w - 1}} \right)' = \left( {{w^2 -
2w + 5} \over {1 -
w}} \right).
\end{equation}
This in turn gives the bounce rule
\begin{equation}
\label{twistcurvbounce}
w' = \left( {{w - 5} \over {w - 1}} \right).
\end{equation}
Note that Eq.~(\ref{twistcurvbounce}) maps the region $1 < w < 3$ into $w <
-1$ so that the combined bounce will always be followed by a kinetic bounce.

The previous analysis must be modified for the combined kinetic-twist
bounce because successive kinetic and twist bounces do not bring $w$ into the
range leading to a curvature bounce. To do this, three bounces are
required---either kinetic-twist-kinetic or twist-kinetic-twist. Either choice
leads to the same rule
\begin{equation}
\label{kintwistbounce}
w' = \left( {{-w + 3} \over {w + 1}} \right) .
\end{equation}
To avoid any implicit assumption regarding the number of bounces which form
the combined bounce, we consider only conserved quantities . The energy
quantity conserved by
$H = H_0 + H_{kin}+H_{twist}$ is
\begin{equation}
\label{twistkinE}
E ={{\pi_P^2} \over {4 \pi_\lambda}}
\,+\,{e^{-2P} \pi_Q^2
\over {4 \pi_\lambda}} \,+\,
\sigma\,\pi_\lambda \,e^{(\lambda + 2P +
3\tau)/2} K^2\, +\, 3\pi_\lambda .
\end{equation}
The second conserved quantity arises in the event that $H_{kin} = H_{twist}$.
In that case, $\pi_P + 2 \pi_\lambda$ is conserved since its time derivative
is proportional to $H_{kin} - H_{twist}$. This leads (using the 
previous procedures) to
\begin{equation}
\label{kintwist2}
\left( {{ w^2\, +\, 3} \over {w \,+\,1}} \right)' = \left( {{
w^2\, +\,3} \over
{w \,+\,1}} \right)
\end{equation}
which has the non-trivial solution Eq.~(\ref{kintwistbounce}). Since it is
unlikely that $H_{kin} = H_{twist}$ for any extended time (this would require
both the growth rates and coefficients to be equal), one would not expect the
generic behavior to include bounces of this type. In fact, none were seen in
the numerical simulations. The bounce rules are summarized in Table
\ref{bounceruletable}.

Using the bounce laws (\ref{kinbounce}), (\ref{curvbounce}), and
(\ref{twistbounce}) and Table~\ref{uval} for the BKL parameter $u$ in
terms of $v = |w|$, the change in $u$ during each type of bounce may be
found. Clearly, Eq.~(\ref{kinbounce}) yields $u' = u$ for any kinetic
bounce, since the change in
sign of $w$ does not change $v$. In a curvature bounce, the initial
range $w \ge 1$ yields two possible relationships between $v$ and $u$
while the final $w'$ can involve all three. The possibilities are shown in
Fig.~1(a). All possibilities yield the standard BKL map for $u$: i.e.~$u' =
u\,-\,1$ if $u \ge 2$ and $u' = 1/(u\,-\,1)$ if $1 \le u \le 2$. A similar
construction for the twist bounce is shown in Fig.~1(b). Here we see that
$u' = u$ results for all initial values of $w$. The rule for $u$ is the same in a twist
bounce, or in a combination twist--kinetic bounce, as
in a kinetic bounce.  Furthermore, since a solution
to any of the three subhamiltonians, $H_0 + H_{kin}$,
$H_0 + H_{twist}$, or $H_0 + H_{kin} + H_{twist}$ is a one
parameter family of Kasner spacetimes, it follows that
the generalized Kasner exponents are (approximately)
constant in time when any one of these subhamiltonians essentially
governs the evolution, and therefore $u$ is (approximately) constant.
The kinetic and twist potentials are each, at any spatial point,
a centrifugal wall.  This means \cite{ryan,jantzen} that the identity
of the principal axis associated with the growing cosmological
scale factor changes during the bounce.  However, the natures
of the bounces are quite different.
During a twist bounce or a combined kinetic twist bounce at a point
in space, the rotation of the principal axes is such that one of
them is orthogonal to the $T^2$ symmetry orbits before the bounce
and tangent after, while another principal direction is tangent
before the bounce and orthogonal to the symmetry orbits after.
During a kinetic bounce the rotation of the principal directions
is in the symmetry plane.  One principal direction is orthogonal
to the symmetry orbit throughout the bounce. In Appendix B
a comparison is made between the dynamics near the singularity
at a spatial point and the dynamics of tilted Bianchi II models
studied in \cite{HBW}.

In Sec.~IV, we shall examine the validity of KEA and the bounce laws in
numerical simulations. However, explicit solutions through the kinetic,
curvature, and twist bounces are known. These may be used to generate
further predictions which can in turn be explored in numerical
simulations. We leave these for future research. The explicit
bounce solutions are given in Appendix A.

\subsection{Exceptional Points}
In a generic $T^2$ symmetric spacetime there are
nongeneric points at which the gravitational field does {\it not}
evolve away from an era of velocity dominated evolution
in the manner that we have described thus far.  There are three
cases in which this happens: \,  $w = 1$ during
a Kasner epoch, $\partial_\theta Q = 0$
during a Kasner epoch with $w > 1$, and
$\pi_Q = 0$ during a Kasner epoch with $w < 0$. In each case, one can
give rough arguments which indicate that bounces 
are likely to occur.
We state these here.

We first consider the case $w = 1$.
The twist bounce solution given in Appendix~A is {\it not}
defined for $w = 1$.  One can, it turns out, write
down a solution generated by $H = H_0 + H_{twist}$ explicitly in
terms of $\tau$ in this case.  It blows up at finite $\tau$.
It is a Kasner spacetime with $\partial_\theta P =0$,
with nonvanishing twist constant, $K$, and with $w = 1$.
However, we claim that this is not a good approximation to
the dynamics at a $w = 1$ exceptional point in a generic spacetime. 
Generally it will not be the case that $\partial_\theta P$
vanishes.  This leads to a situation in which the
exponential factors do not control the terms they appear in.
In the solution that blows up, $\pi_\lambda \rightarrow 0$.
But if $\pi_\lambda$ gets small enough,
$1 \over 4 \pi_\lambda$
wrests control of $H_{small}$ from $e^{-2 \tau}$ and $H_{small}$ 
becomes relevant.  The subhamiltonian that governs the evolution
in this case is $H = H_0 + H_{small} + H_{twist}$. This has (after
a canonical transformation) the same structure as the Hamiltonian for
polarized magnetic Bianchi VI$_0$, in which case there are rigorous
results which show that the solution does not blow up in finite time,
and which predict the bounce rule. One also notes
that the $\tau$ dependence of the argument of the exponential in
$H_{curv}$ vanishes if
$w = 1$. This means that, generically, $H_{curv}$ contributes a constant (not
an exponentially decaying) term to Eq.~(\ref{pipev}) for $\partial_\tau
\pi_P$.
But $w = 1$ yields
$\gamma := \pi_P - 2\pi_\lambda = 0$ for the quantity which is conserved
in
the twist bounce. The change in $\gamma$ due to the term from $H_{curv}$
will remove the exceptional point condition.  

We next consider the case that $\partial_\theta Q$
crosses zero at some $\theta_0$ during a Kasner
epoch with $w > 1$. The curvature bounce is suppressed in a neighborhood
of $\theta_0$.  The closer to the exceptional point, the longer the bounce
is suppressed.  If $w < 2$ there will be a twist bounce
which sends $w \rightarrow w' > 3$.  As the
neighborhood on which the curvature bounce
is suppressed gets
smaller and smaller, $\partial_{\theta \theta} P$
and $\pi_Q$ grow exponentially if $w > 2$
\cite{gowdyphem}
(as can be seen in the
curvature bounce solution~(\ref{expcurv}) by considering
the case that $\zeta$ crosses zero).  This wrests control
of $H_{kin}$ and terms in the evolution equations
derived from $H_{curv}$ from the exponential factors,
and causes $w$ to decrease until it is again
less than two, in which case the twist potential
begins to grow again, so bounces will continue.

In the case that $\pi_Q$ crosses zero
during a Kasner epoch with $w<0$ a similar
mechanism (here $\partial_{\theta \theta} P$
and $\partial_\theta Q$ grow exponentially
if $w < -1$) (\cite{gowdyphem} and see Appendix~A)
causes $w$ to become greater than $-1$,
in which case the twist potential
starts growing, so bounces will continue. 

These arguments suggest  that there continue to be oscillations even at
or near exceptional points. The behavior of the gravitational field at an
exceptional point is delicate, however. It may, for example, be the case
that higher order terms play an important role. Further study of
exceptional points is needed, and is now underway.

\subsection{Minisuperspace picture and twist bounces}
One of the more useful (and most pictorial) ways to study the dynamics of
spatially homogeneous cosmological solutions of Einstein's equations is via
the minisuperspace (MSS) picture. We would like to relate our discussion thus
far of local oscillatory behavior and bounces in $T^2$ symmetric solutions to
the MSS approach. In particular, we wish to see how the local twist bounces
appear in the MSS picture.

We recall that the MSS approach represents spatially homogeneous spacetimes
as follows:  For each choice of the spatial isometry group $G$ (e.g.,
$G={\cal R}^3\leftrightarrow $ Bianchi I, $G = SU(2) \leftrightarrow$ Bianchi
IX) one chooses a fixed group invariant frame, and then one can parametrize
the set of 3-geometries invariant under $G$ using the MSS variables $\Omega$
(volume), $\beta_\pm$ (anisotropy) and $\mu_i$ (if the metric is not
diagonal). Using a simple choice of lapse and shift, one can represent a
spacetime by a trajectory in the MSS configuration space $(\Omega(t),
\beta_+(t),\beta_-(t))$, with the $\mu_i(t)$ playing a subsidiary role. It
follows from Einstein's equations, adapted to the spatial homogeneity, that
the trajectories corresponding to solutions evolve via Hamilton's equations,
with the Hamiltonian potential proportional to the scalar curvature
${}^3\negthinspace R(\beta_+,\beta_-,\Omega)$.

For Bianchi I spacetimes, ${}^3\negthinspace R = 0$, so that the
trajectories are straight lines. For the diagonal Bianchi IX solutions, the
potential may be represented by triangular walls, as in Fig.~2. (Note that,
in this figure, the dynamics has been projected onto the anisotropy plane
and rescaled $( \beta_+/|\Omega|, \beta_-/|\Omega|)$ so that the location
of the potential walls is independent of $\Omega$). The dynamics then
consists of straight line segments (effectively Kasner intervals)
punctuated by intermittent reflections off one or the other of the walls
(these are the bounces). The Kasner map describes the sequences of bounces,
and relates the parameters of the Kasner interval after the bounce to those
of the Kasner interval before the bounce. The MCP prediction that an
infinite sequence of bounces should occur follows from the closed nature of
the region in MSS configuration space which is bounded by the walls. 

Before tying this picture to the local behavior of $T^2$ symmetric
solutions, we note what happens to the MSS description of Bianchi IX
spacetimes if the metric is not diagonal. In that case, additional
(``centrifugal'') walls appear, bisecting the angles of the triangle. One
or more may be present, depending on the orientation of the rotational axis
relative to the principal axes in the spacetime. The centrifugal walls do
affect the bounces; however, a very slightly modified Kasner maps allows one
to predict the effect of these bounces on the Kasner interval transitions.

To obtain a MSS type picture for the local dynamics of $T^2$ symmetric
solutions, it is useful to first do so for certain subfamilies. For the Gowdy
subfamily (twist equal to zero), we compare the Bianchi I spatial metric
\begin{equation}
\label{type1metric}
\gamma_I = e^{2\Omega-4\beta_+} d\theta^2 +
e^{2\Omega+2\beta_++2\sqrt{3}\beta_-} dx^2+
e^{2\Omega+2\beta_+-2\sqrt{3}\beta_-} dy^2
\end{equation}
with the polarized (and diagonal) Gowdy spatial metric
\begin{equation}
\label{gpolmetric}
\gamma_{PG} = e^{(\lambda + \tau)/2} d\theta^2 + e^{-\tau + P} dx^2 + e^{-\tau
- P} dy^2
\end{equation}
to find the following relations between the MSS variables
$(\Omega,\beta_+,\beta_-)$ and the Gowdy metric components
\begin{equation}
\label{mss}
P= 2\sqrt{3}\beta_- \quad, \quad \lambda = 6(\Omega - \beta_+) \quad,
\quad \tau = -2 (\Omega + \beta_+).
\end{equation}
(These identifications are not unique, since we have singled out one
direction, $\partial/\partial \theta$, in the Bianchi I spacetime to identify
with the direction of spatial dependence in the Gowdy spacetime.)
Then if we rotate in the $x$-$y$ plane by an angle $\xi$ and make
identifications, $\tau$ and $\lambda$ are unchanged but we have
\begin{equation}
\label{prot}
e^P = e^{2\sqrt{3}\beta_-} \cos^2\xi + e^{-2\sqrt{3}\beta_-} \sin^2\xi
\end{equation}
and
\begin{equation}
\label{qrot}
Q = {{\sin \xi \cos \xi (e^{2\Omega + 2\beta_++2\sqrt{3}\beta_-} - e^{2
\Omega -4 \beta_+})} \over {e^{2
\Omega -4 \beta_+} \cos^2 \xi + e^{2\Omega + 2\beta_++2\sqrt{3}\beta_-}
\sin^2 \xi}} \ .
\end{equation}
Now adding spatial dependence on $\theta$ to the Kasner solution obtained by
the rotation through
$\xi$ yields a generic Gowdy spacetime where we recall that the Gowdy
spacetimes may be obtained from $T^2$ symmetric spacetimes by setting the
twist constant $\kappa$ to zero and $\pi_\lambda$ to $1/2$. Unpolarized Gowdy
spacetimes have two non-vanishing potentials, $H_{kin}$ and $H_{curv}$
(specialized to the Gowdy case). It follows from (\ref{prot})
that the Gowdy potential $H_{curv}$ will be of order unity on the lines
labeled $C$ and $C'$ in Fig.~2 if that diagram is assumed to depict a
local MSS for a Gowdy spacetime. Similarly, $H_{kin}$ will be of order
unity on the line labeled $K$ in Fig.~2 when the distances from the
system point to $C$ and $C'$ are equal.

In a similar fashion, a rotation through an angle $\omega(\tau)$ in the
$\theta$-$y$ plane leads to new identifications which correspond to
those appropriate to a polarized $T^2$ symmetric spacetime. The
identifications are based on the metric (\ref{metric}) with $Q$ and $G_1$
set to zero in the spatial metric. We find
\begin{equation}
\label{expm2p}
e^{-2P} = e^{-4\sqrt{3}\beta_-}\cos^2 \omega + e^{-2\sqrt{3}\beta_- -
6\beta_+} \sin^2 \omega,
\end{equation}
\begin{equation}
e^{-2\tau} = e^{4\beta_+ + 4 \Omega} \cos^2 \omega + e^{2 \sqrt{3}
\beta_- - 2 \beta_+ + 4\Omega} \sin^2 \omega,
\end{equation}
and the twist potential combination
\begin{equation}
\label{centtwist}
e^{(\lambda + 2P + 3\tau)/2} = \left( e^{3 \beta_+ - \sqrt{3} \beta_-}
\cos^2 \omega + e^{-3 \beta_+ + \sqrt{3} \beta_-} \sin^2 \omega
\right)^{-1}.
\end{equation}
Each of these identifications contains two exponential terms on the right
hand side, only one of which can be large at a given time. If we choose one of
these to represent the centrifugal wall (e.g. the one consistent with the
original Kasner identifications for $P$ and $\tau$), we find that the
centrifugal wall is of order unity along the line labeled $T$ in Fig.~2.

We claim that in the general $T^2$ symmetric models, the walls $C$,
$C'$, $K$, $T$, and the walls allowed by alternate identifications (e.g.
$T'$ in Fig.~2) are present in the local MSS picture. In that case, the local
dynamics is confined to the shaded region in Fig.~2. A twist
producing rotation yields a term
$p_\omega^2/\sinh^2(6\beta_+ - 2 \sqrt{3} \beta_-)$ in the Kasner or
mixmaster Hamiltonian \cite{ryan}. Identification of the exponentials from
the metric is not the same as constructing the relevant
potential---hence, the absence of a minus sign in the denominator of
(\ref{centtwist}). However, if only one of the two exponentials in the
denominator is large, such a sign cannot be detected. Presumably, an
analysis such as that in \cite{sign} should demonstrate the equivalence
between the twist potential and the Kasner centrifugal potential in the local
MSS picture.

\section{Numerical Results}
MCP arguments give a qualitative prediction of what behavior
one might see in solutions of Einstein's equations. They are, however,
neither rigorous nor complete. Thus it is important to compare the
solution behavior predicted by MCP arguments with the behavior observed
in numerical studies of $T^2$ symmetric spacetimes. As we discuss here,
the agreement is remarkable. 

Eqs.~(\ref{pev})--(\ref{pilamev}) are solved
numerically using a second order Iterative Cranck-Nicholson (ICN) method
(see for example \cite{teukolsky}). Symplectic methods used in our previous
studies
\cite{numinv,bgimw} fail for these models. Apparently, the operator splitting
used in the symplectic algorithm allows a pathological behavior in
$\pi_\lambda$ which is suppressed by other (more standard) methods. The
CN algorithm can be shown to be more stable but less accurate than
symplectic methods prior to blow-up of the latter. 

In our numerical studies, we have examined the evolution of the
gravitational field for a wide variety of sets of initial data. We get
qualitatively similar results in all cases. The graphical results
displayed and discussed here are primarily based on the representative
choice of initial data with $P =0,\ \pi_P = 5 \cos (\theta + \pi/5),\ Q = \cos (\theta + \pi/5),
\ \pi_Q = 0,\ \lambda = 0, \ \pi_\lambda = 1/2$ and with $\sigma = 1$ and
$\kappa = 10^{-4}$. This particular choice of data is useful in that it
leads to early onset of twist bounces.

A crucial feature of BKL (or AVTD) behavior is that the evolving fields
satisfy the KEA within finite time at each spatial point $\theta$. To
check for the onset (and later, the recurrence) of field evolution
consistent with the KEA, we monitor $w = \pi_P/(2\pi_\lambda)$ at every
spatial point. Typical behavior (at three spactial points) is shown in
Fig.~3.
The analysis
of Sec. III predicts that $w(\theta)$ should be asymptotically
piecewise constant. This is seen to be the case. Careful examination (in
Fig.~4) shows that the constancy of $w(\theta)$ becomes an ever better
approximation as $\tau \to \infty$.

To study the validity of the MCP predictions for the change of the value
of $w(\theta)$ following a bounce, given its value before
[Eqs.~(\ref{kinbounce}), (\ref{curvbounce}), (\ref{twistbounce}),
(\ref{twistcurvbounce}), and (\ref{kintwistbounce})], we measure the $w$
values in all the Kasner epochs. The next value is then predicted using
all possible bounce laws. If the new value of $w$ obeys any one of these,
then the difference
$|\Delta w|$ between the actual and predicted values should be much
smaller
than those obtained by chance. For the indicated initial data, at a
typical spatial point, the first twist bounce is followed by a sequence
of alternating kinetic and curvature bounces. Typically, the comparison
of actual to bounce law predictions improves as $\tau$ increases.
Typical
data are shown in Fig.~5. At some spatial points, a second twist bounce
occurs. Fig.~6 shows the values of $|\Delta w|$ computed for all bounces
using the twist bounce law. The small values are indicative of the
actual
twist bounces. Fig.~7 shows the behavior of $P$ at spatial
points with a second twist bounce having $w > 1$ and $w < 1$. 

Interestingly, we find in our numerical simulations a number of spatial
points where a combined curvature-twist bounce has occured. This is shown
in the plot of combined
bounce rule $|\Delta w|$s shown in Fig.~8. Fig.~9 shows $w(\tau)$ at
such a spatial point with the same quantity at nearby spatial points. At
the latter, it is clear that the combined bounce has split into separate
curvature and twist bounces.

As illustrated in Fig.~10, for small values of $w$ after the first twist
bounce, the evolution changes very little if we change the spatial
resolution. This indicates convergence (in the computational analysis
sense) at such spatial points. If $w$ is large after the first twist
bounce, then the evolution does appear to be somewhat spatial-resolution
dependent (as is also illustrated in Fig.~10). This reflects some
sensitivity to initial conditions at the first twist bounce where a
subtraction appears in the denominator of the bounce law. However, within
a given trajectory, the agreement with bounce laws is more convergent
with increasing spatial resolution as seen
in Fig.~11. The waveforms at $ \tau = 61.66 $ are shown for different
spatial resolutions in Fig.~12. Inclusion of adaptive mesh refinement in
these codes is in progress
\cite{bbdg}.

In Fig.~13, a section of the spatial axis is shown for graphs of
$\partial_\theta Q$, $\partial_\theta P$, and $\pi_Q$ vs $\theta$ at a
late value of $\tau$. If any of these quantities vanish at some
$\theta_0$, exceptional behavior results there. The zero crossings are
shown. For the given initial data, the density of exceptional points
increases rapidly with time as is shown in Fig.~14. Since such points
are generated by evolving small scale spatial structure, one could argue
that exceptional points should become a dense set (of measure zero) as
$\tau \to \infty$. This means that any rigorous statements about the
nature of these solutions must include consideration of the behavior at
exceptional
points.

At a few points, there are anomalies where the bounce laws are violated.
This seems to be a consequence of inadequate resolution since the effect
disappears at higher spatial resolution. The data for $P$, $Q$, and
$\lambda$ (at two spatial resolutions) for the results discussed in this
section are shown in Fig.~15.

The simulations illustrated here cannot be run significantly beyond
$\tau \approx 75$ (as indicated on various figures). The reason becomes
clear from examination of the values of $w$ at $\tau \approx 75$. At
several values of
$\theta$, $w(\theta)$ is very close to (and less than) unity. From
Eq.~(\ref{twistbounce}), it is then clear that the next bounce should be a
twist bounce with a very large new value of $w$. But $w >> 1$ 
produces numerical overflows in at least one exponential term (depending on
the sign
of $w'$) in the equations of motion (\ref{pev})--(\ref{pilamev}). This
is
a ``physical,'' resolution-independent numerical instability. While possible
ways to resolve this problem involve checking the value of $w(\theta)$ (and
thus slowing the code), there is no difficulty in principle in either
using arbitrary precision arithmetic (see for example \cite{press}) or the MCP
solution (see Appendix A) for the next kinetic or curvature bounce. Fig.~16 is
constructed by using the twist bounce rule on the computed array
$w(\theta,\tau)$ to obtain $w'(\theta,\tau)$. Momentary (pointlike)
large values of
$|w|$ arise during bounces when the KEA does not hold. The 
persistent large values of $|w'|$ for $\tau \approx 75$ indicate that
dangerously large values are likely to arise after the next bounce at some
spatial points.

\section{Conclusions}
We have examined the approach to the singularity in $T^2$ symmetric
vacuum spacetimes. Numerical simulation provides strong support for
the contention that these models reach an asymptotic regime where the KEA
holds. Given the KEA, we then predict, using the MCP, the rules relating
one Kasner epoch to the next. Again the numerical simulations show
remarkable agreement with the MCP predictions. 

These spacetimes may be understood as another example family whose members
exhibit local mixmaster dynamics in the vicinity of the singularity. Yet
the local mixmaster dynamics shown here differs from that studied in
magnetic Gowdy models. In that case, the local MSS potential is closed
by a magnetic wall which replaces one of the curvature walls that one
would expect in a locally Bianchi IX spacetime. In $T^2$ symmetric
spacetimes, the essentially non-diagonal centrifugal wall closes the
potential. 

Several questions remain open. First, we may ask whether there are an
infinite number of bounces. In the absence of exceptional points, one
could start from any value of $w$ and apply the bounce rules
(\ref{kinbounce})--(\ref{twistbounce}) indefinitely. We have argued that
the most common exceptional points with $\partial_\theta P = 0$,
$\partial_\theta Q = 0$, or $\pi_Q = 0$ do not cause the bounces to
terminate. However, we have seen that the number of exceptional points
increases as ever smaller scale spatial structure is produced by bounces
which occur at different places at different times. Any rigorous
discussion of the asymptotic behavior of $T^2$ symmetric models must deal
with the exceptional points. We do not yet know the role, if any, played
by exceptional points where higher derivatives also vanish. Detailed
discussions of exceptional points will be given elsewhere \cite{longmaggow}.

A second open question concerns the relationship between $T^2$ symmetric
and $U(1)$ symmetric models. If a {\it diagonal} Bianchi IX metric is
expressed in terms of the $U(1)$ variables, all features observed up to
now in generic $U(1)$ models \cite{beverly-vinceU(1)} may be explained in
terms of local mixmaster dynamics \cite{sign}. It is not yet known whether
analogs of the twist bounce (a feature of {\it non-diagonal} Bianchi IX
models) have been missed or suppressed in the existing $U(1)$ simulations. A
detailed discussion of the relationship between the two classes of spacetimes
will be given elsewhere \cite{t2asu1}.

Even with these open questions, we have provided strong support for the
validity for $T^2$ symmetric spacetimes of the BKL picture in its most
general (local, non-diagonal Bianchi IX) form. We have also provided yet
another example of the power of the MCP in the analysis of the approach to
the singularity in inhomogeneous cosmologies. Finally, we have shown how
this class of spacetimes allows accurate numerical simulations yet
provides a highly non-trivial manifestation of local mixmaster dynamics.

\section*{Acknowledgements}
B.K.B. and J.I. would like to thank the Albert Einstein Institute (Golm)
for hospitality. B.K.B. would like to thank the Institute for
Geophysics and Planetary Physics at Lawrence Livermore National
Laboratory for hospitality. M.W. would like to thank Alan Rendall for
insightful remarks. This work was supported in part by National Science
Foundation Grants PHY9800103 and PHY9800732. Some of the numerical
simulations discussed here were performed at the National Center for
Supercomputing Applications of the University of Illinois.

\section*{Appendix A \, \, Explicit Solutions for the Bounces}
If the gravitational field at $(\theta_0,\tau_{(\theta_0)})$
satisfies the KEA, then the local evolution of the gravitational
field quantities $\{P,Q,\mu,\lambda,\pi_P,\pi_Q,\pi_\lambda\}$
takes the simple velocity dominated form,
but at nonexceptional points at
least one of the Hamiltonian potentials grows in time.
To understand what happens as one of the potentials
becomes significant, we study the evolution
of the fields for each of the three governing
Hamiltonian densities $H=H_0 + H_{kin}$,
$H=H_0 + H_{curv}$ and $H=H_0 + H_{twist}$.
Letting $\{\hat{P},\hat{Q},\hat{\mu},\hat{\lambda},\hat{\pi}_P,
\hat{\pi}_Q,\hat{\pi}_\lambda\}$ denote the data values
at $(\theta_0,\tau_0)$, with $\tau_0 = \tau_{(\theta_0)}$,
we obtain in each case the explicit\footnote{In the case of
the twist bounce, we use an implicitly defined function.} 
general solution on a neighborhood of $\theta_0$.

\subsection*{A.1. Kinetic Bounce}
\begin{eqnarray}
\label{appkinham}
H & = & H_0 + H_{kin} \nonumber \\
& = & {1 \over 4 \pi_\lambda } (\pi_P^2 + e^{-2 P} \pi_Q^2).
\end{eqnarray}

For a kinetic bounce to occur for some $\tau > \tau_0$,
we need $\hat{\pi}_Q \neq 0$ and we need $\hat{w} < 0$, or
equivalently, $\hat{\pi}_P < 0$.  We presume that both
of these conditions hold for the data at $(\theta_0,\tau_0)$.

Now let us define the following series of convenient
constants\footnote{These constants depend on $\theta$,
and the solution will be a good approximation to
the evolution through the bounce on a neighborhood
of $(\theta_0)$.  Knowledge of the solution on
a spatial neighborhood is necessary to confirm
that the exponential factors generically do, both
before and after the bounce (and in the case of
the neglected potentials also during the bounce), 
control the terms in which they appear.
It is also necessary for analysis of the
exceptional points, since the spatial derivatives
play a crucial role.  But we are discussing here
field evolution in $\tau$ at the fixed point
$\theta_0$, so we write the quantities as
functions of time alone.} (all depending
upon the data at $(\theta_0,\tau_0)$.
\begin{eqnarray}
\beta & := & {1 \over 2 \hat{\pi}_\lambda}
[\hat{\pi}_P^2 + e^{-2 \hat{P}} \hat{\pi}_Q^2]^{1\over2}, \\
\zeta & := & {e^{-\hat{P} - \beta \tau_0} \hat{\pi}_Q \over
\hat{\pi}_P - 2 \beta \hat{\pi}_\lambda}, \\
\alpha & := & {e^{\hat{P} + \beta \tau_0} \over
1 + \zeta^2 e^{2 \beta \tau_0} }.
\end{eqnarray}
We note that the KEA, with $\hat{\pi}_P < 0$,
implies that $\beta \approx -\hat{\pi}_P / 2 \hat{\pi}_\lambda
= - w$.  We also note that the KEA implies
that $\zeta$ is very small, and $\hat{\pi}_Q = 0 \Leftrightarrow
\zeta = 0$.

In terms of these constants, the solution governed by the
density~(\ref{appkinham}) and matching the initial conditions
at $\tau_0$ is (for fixed $\theta_0$)
\cite{gowdyphem,numstudy,numinv}
\begin{eqnarray}
\label{expkin}
\nonumber
\end{eqnarray}
\vspace*{-\baselineskip}
\vspace*{-\baselineskip}
\vspace*{-\baselineskip}
\begin{eqnarray}
\label{expkinp}
P(\tau) & = & \hat{P} - \beta \, (\tau - \tau_0) +
\ln \left( {1 + \zeta^2 e^{2 \beta \tau} \over
1 + \zeta^2 e^{2 \beta \tau_0}} \right), \\
\label{expkinq}
Q(\tau) & = & \hat{Q} + \zeta \, e^{-\hat{P} + \beta \tau_0}
-{\zeta \, e^{2 \beta \tau} \over \alpha ( 1 + \zeta^2
e^{2 \beta \tau})} ,\\
\lambda(\tau) & = & \hat{\lambda} - \beta^2 (\tau - \tau_0), \\
\pi_P(\tau) & = & -2 \, \hat{\pi}_\lambda \, \beta \left( { 1 -
\zeta^2 e^{2 \beta \tau} \over 1 + \zeta^2 e^{2 \beta \tau}}
\right) ,\\
\pi_Q(\tau) & = & \hat{\pi}_Q, \\
\pi_\lambda(\tau) & = & \hat{\pi}_\lambda.
\end{eqnarray}

\subsection*{A.2. Curvature Bounce}
\begin{eqnarray}
\label{appcurvham}
H & = & H_0 + H_{curv} \nonumber \\
& = & {1 \over 4 \pi_\lambda } (\pi_P^2 + e^{2 (P - \tau)}
(\partial_\theta Q)^2).
\end{eqnarray}
In this Hamiltonian, a spatial derivative term appears.
However, since $\pi_Q$ does not appear,
the general solution for fields governed by~(\ref{appcurvham})
is relatively straightforward to derive.  This
Hamiltonian is in fact related to~(\ref{appkinham})
by a canonical transformation.
The similarity in structure can be seen in the two explicit
solutions, but we do not discuss this further here.

For a curvature bounce to occur for some $\tau > \tau_0$,
we need $\partial_\theta Q \neq 0$ and we need $\hat{w} < 1$, or
equivalently, $\hat{\pi}_P < 2 \hat{\pi}_\lambda$.
We presume that both of these conditions hold for the
specified data at $(\theta_0,\tau_0)$.

To write out explicitly the general solution for $H$
in~(\ref{appcurvham}), it is again useful to first
define a set of constants (depending on the data
at $\tau_0$):
\begin{eqnarray}
\tilde{\beta} & := & { 1 \over 2 \hat{\pi}_\lambda }
[(2 \hat{\pi}_\lambda - \hat{\pi}_P)^2 +
e^{2 ( \hat{P} - \tau_0 )}
(\partial_\theta \hat{Q})^2]^{1\over2} ,\\
\tilde{\zeta} & := & -{e^{\hat{P} - (1 + \tilde{\beta}) \tau_0}
\partial_\theta Q \over \hat{\pi}_P -
2 \hat{\pi}_\lambda} ( 1 - \tilde{\beta}), \\
\tilde{\alpha} & := & {e^{-\hat{P} +
(\tilde{\beta} - 1) \tau_0} \over
1 + \tilde{\zeta}^2 e^{2 \tilde{\beta} \tau_0} }.
\end{eqnarray}
We note that the KEA, with
$2 \hat{\pi}_\lambda - \hat{\pi}_P < 0$
implies that $\beta \approx -(2 \hat{\pi}_\lambda - \hat{\pi}_P)
 / 2 \hat{\pi}_\lambda = (w-1)$ and that $\tilde{\zeta}$ is
very small.  $\partial_\theta \hat{Q} = 0 \Leftrightarrow
\tilde{\zeta} = 0$.
The solution takes the following form \cite{thesis}:

\begin{eqnarray}
\label{expcurv}
\nonumber
\end{eqnarray}
\vspace*{-\baselineskip}
\vspace*{-\baselineskip}
\vspace*{-\baselineskip}
\begin{eqnarray}
\label{expcurvp}
P(\tau) & = & \hat{P} + ( 1 + \tilde{\beta}) \, (\tau - \tau_0) -
\ln \left( {1 + \tilde{\zeta}^2 e^{2 \tilde{\beta} \tau} \over
1 + \tilde{\zeta}^2 e^{2 \tilde{\beta} \tau_0}} \right), \\
\label{expcurvq}
Q(\tau) & = & \hat{Q}, \\
\lambda(\tau) & = & \hat{\lambda} -
(1 + \tilde{\beta})^2 (\tau - \tau_0)
+ 2 \ln \left( {1 + \tilde{\zeta}^2 e^{2 \tilde{\beta} \tau} \over
1 + \tilde{\zeta}^2 e^{2 \tilde{\beta} \tau_0}} \right) ,\\
\pi_P(\tau) & = & 2 \, \hat{\pi}_\lambda + 2 \, \hat{\pi}_\lambda
\, \tilde{\beta} \left( { 1 -
\tilde{\zeta}^2 e^{2 \tilde{\beta} \tau}
\over 1 + \tilde{\zeta}^2 e^{2 \tilde{\beta} \tau}}
\right) ,\\
\label{expcurvpiq}
\pi_Q(\tau) & = & \hat{\pi}_Q -
e^{2 (\hat{P} - \tau_0)}
\left\{(\partial_\theta \tilde{\alpha}) \tilde{\zeta}
- \tilde{\alpha} (\partial_\theta \tilde{\zeta})
- 2 \tilde{\alpha} (\partial_\theta \tilde{\beta})
\tilde{\zeta} \tau_0 + \tilde{\zeta}^2 e^{2 \tilde{\beta} \tau_0}
[(\partial_\theta \tilde{\alpha}) \tilde{\zeta} +
\tilde{\alpha} (\partial_\theta \tilde{\zeta})] \right\}
\nonumber \\
& & \; \; \; \; \; \; e^{2 (P - \tau)}
\left\{(\partial_\theta \tilde{\alpha}) \tilde{\zeta}
- \tilde{\alpha} (\partial_\theta \tilde{\zeta})
- 2 \tilde{\alpha} (\partial_\theta \tilde{\beta})
\tilde{\zeta} \tau + \tilde{\zeta}^2 e^{2 \tilde{\beta} \tau}
[(\partial_\theta \tilde{\alpha}) \tilde{\zeta} +
\tilde{\alpha} (\partial_\theta \tilde{\zeta})] \right\} ,\\
\pi_\lambda(\tau) & = & \hat{\pi}_\lambda .
\end{eqnarray}

Note the appearance of $P(\tau)$ in the exponent
of~(\ref{expcurvpiq}); one may substitute in $P(\tau)$
from~(\ref{expcurvp}) to obtain an expression in
terms of the initial data.

\subsection*{A.3. Twist Bounce}
\begin{eqnarray}
\label{apptwistham}
H & = & H_0 + H_{twist} \nonumber \\
& = & {1 \over 4 \pi_\lambda } \pi_P^2 + \sigma \,
\pi_\lambda e^{\lambda + 2 P + 3 \tau \over 2 } K^2.
\end{eqnarray}

For a twist bounce to occur for some $\tau > \tau_0$,
we need $-1 < \hat{w} < 3$; that is,
$-1 < \hat{\pi}_P / 2 \hat{\pi}_\lambda < 3$.
We presume this condition holds.
Note that, by assumption, $K \neq 0$; and note that it
follows from the constraint~(\ref{t2mu}) that
$\pi_\lambda > 0$. Hence $H_{twist}$ is always positive.
While we can find the solution to~(\ref{appcurvham})
for all values of initial data with $w \in (-1,3)$,
one finds that if $w = +1$, the solution blows
up in finite time \cite{longmaggow}.
(See Sec. III.) We thus
presume that $w \in (-1,1) \cup (1,3)$.

As for the other two bounces, in writing out
the solution for the twist bounce it is useful
to define a few constants (depending on
the initial data):\footnote{The twist bounce
solution can be obtained by a canonical transformation
of the magnetic bounce solution in the magnetic
Gowdy case.  That solution is given in
the context of magnetic Bianchi VI$_0$ in \cite{newalg}.}
\begin{eqnarray}
\xi & := & \hat{H}_0 + \hat{H}_{twist} -\hat{\pi}_P
+ 5 \hat{\pi_\lambda} ,\\
\phi & := & \hat{H}_{twist} ,\\
\nu & := & [\xi^2 - 4 ( \hat{\pi}_P -
2 \hat{\pi}_\lambda)^2]^{1 \over 2}.
\end{eqnarray}
We also find it necessary to work with an implicitly
define time: we define the variable $T$ via the
implicit equation
\begin{equation}
\label{implicittime}
\tau(T) - \tau_0 = -{1 \over 2} \ln[\cosh(\nu T) -
{8 \hat{\pi}_\lambda - \xi \over \nu} \sinh(\nu T)]
+ {\xi T \over 2}.
\end{equation}
Note that for the case we are excluding, $\hat{w}=1$,
$\tau(T) - \tau_0 = 0$.  But if $\hat{w} \neq 1$,
we verify that $\tau(T)$ is smooth
and is a strictly increasing
function of $T$,
$${d\tau \over dT} = 4 \pi_\lambda.$$
Hence, it is invertible, and we can solve for
$T(\tau)$ in principle.  (In the numerical
calculations which use the symplectic algorithm
we use a rootfinder to find $T(\tau)$.)

If we define
\begin{equation}
\label{defs2}
S(T) := { \sinh (\nu T) \over \nu \cosh (\nu T) -
(8 \hat{\pi}_\lambda - \xi)  \sinh (\nu T) },
\end{equation}
the expressions for the solution are the following \cite{thesis}
\begin{eqnarray}
\label{exptwist}
\nonumber
\end{eqnarray}
\vspace*{-\baselineskip}
\vspace*{-\baselineskip}
\vspace*{-\baselineskip}
\begin{eqnarray}
\label{exptwistp}
P(\tau) & = & \hat{P} + \tau - \tau_0 + (\hat{\pi}_P - 2
\hat{\pi}_\lambda) \, T(\tau) ,\\
\label{exptwistq}
Q(\tau) & = & \hat{Q} ,\\
\lambda(\tau) & = & \hat{\lambda} -
2 \ln[ \cosh (\nu T) -   
(8 \hat{\pi}_\lambda - \xi)  \sinh (\nu T) / \nu] \nonumber, \\
& &
-4 \ln [ 1 -2 \phi S(T) ] 
+ ( 2 \hat{\pi}_\lambda - \hat{\pi}_P
- 2 \xi) T \\
\pi_P(\tau) & = & \hat{\pi}_P + 4
\hat{\pi}_\lambda \phi S(T) ,\\
\label{exptwistpiq}
\pi_Q(\tau) & = & \hat{\pi}_Q ,\\
\pi_\lambda(\tau) & = & \hat{\pi}_\lambda ( 1 - 2 \phi S(T)).
\end{eqnarray}

%

\section*{Appendix B \, \, Generalized Kasner Exponents}
When the quantity $\epsilon = -\partial_\tau \mu/2$ is small
the generalized Kasner exponents are approximately
\begin{equation}
\label{exactGowdyGKE}
\kappa_1 \approx {\partial_\tau{\lambda} - 2 \epsilon+
1 \over \partial_\tau{\lambda} - 2 \epsilon-3},
\; \; \; \; \; \;\kappa_2 \approx {2\,(v-1)
\over \partial_\tau{\lambda}-2 \epsilon-3},
\; \; \; \; \; \;\kappa_3 \approx { -2\,(v+1 )
\over \partial_\tau{\lambda}-2\epsilon-3}.
\end{equation}
It follows directly from the evolution equations that
the denominator appearing in Eqs.~(\ref{exactGowdyGKE}) is
bounded above by -3.  The $\kappa$'s are exactly the generalized
Kasner exponents if the twist constant $K=0$.  To derive
Eqs.~(\ref{exactGowdyGKE}), we first define the orthonormal spatial frame,
\begin{eqnarray}
\label{orth1a}
U_1 & = & e^{-(\lambda + \mu + \tau)/4} \,
(\partial_\theta - G_1 \, \partial_x - G_2 \, \partial_y ),\\
U_2 & = & {e^{(-P+\tau )/ 2} \over \sqrt{\sigma}} \, \partial_x, \\
\label{orth1c}
U_3 & = & {e^{(P+\tau )/ 2} \over \sqrt{\sigma}}
\, (\partial_y -Q \partial_x).
\end{eqnarray}
The components of the extrinsic curvature in this frame are
\begin{equation}
\label{extcurv1}
k_{ab} = {1 \over 2}{e^{(-\lambda + 3 \tau)/4} }
\left( \begin{array}{ccc}
{(\partial_\tau \lambda- 2 \epsilon+1) / 2}
& 0  & -\sqrt{\epsilon}\\
0 & w-1 & e^P \partial_\tau Q \\
-\sqrt{\epsilon} &\ \  e^P \partial_\tau Q & \ \ -w-1
\end{array} \right)
\end{equation}
In this frame the twist bounce and kinetic bounce both occur as
bounces off centrifugal potentials.  It is convenient to compare
the oscillatory dynamics of the $T^2$ symmetric spacetimes to that
of the tilted Bianchi II models studied in \cite{HBW} using this
frame, because the form of the extrinsic curvature is the
same in the two cases, and the off-diagonal components of
(\ref{extcurv1}) are significant during a twist bounce and
a kinetic bounce, in turn.  The spatially homogeneous models
studied in \cite{HBW} have a tilted perfect fluid as source.  Note
that, without the source, the constraints rule out the possibility
of oscillatory dynamics in those models, while in the spatially
inhomogeneous $T^2$ symmetric spacetimes similar dynamics are
obtained in vacuum.  Figure II in \cite{HBW} depicts the oscillatory
dynamics which we see at a generic spatial point.  Figure~II~i) in
that reference depicts the curvature bounce solutions (in our
language).  Figure~II~ii) depicts the kinetic bounce solutions
and Figure~II~iii) depicts the twist bounce solutions.  The
identification is fixed by setting $w = 0$ at the point $Q_3$
in their figure, with $w$ increasing in the clockwise direction
around the Kasner circle ($w \to \pm \infty$ at the point $T_3$).
The authors of \cite{HBW} consider an orthonormal frame $e_a$, and the
variables
$(\Sigma_+,\Sigma_-)$.  Let $e_1 = U_2$, $e_2 = U_3$ and
$e_3 = U_1$.  While the conditions placed in \cite{HBW} on the spatial frame
are not satisfied here, they are approximately satisfied at generic   
spatial points near the singularity. Setting
\begin{equation}
\Sigma_+ = {1 \over 2} \, 
\left({ 6 w - \partial_\tau \lambda
+ 2 \epsilon - 3 \over - \partial_\tau \lambda
+ 2 \epsilon +3} \right), \hspace{60pt}
\Sigma_- = {\sqrt{3} \over 2} \, 
\left({ 2 w + \partial_\tau \lambda
- 2 \epsilon + 3 \over - \partial_\tau \lambda
+ 2 \epsilon +3} \right),
\end{equation}
we obtain figure II from \cite{HBW} by noting that
$\Sigma_-/(2 - \Sigma_+)$ is constant in a curvature
bounce solution, $\Sigma_+ - \sqrt{3} \Sigma_-$ is
constant in a kinetic bounce solution, and
$\Sigma_+$ is constant in a twist bounce solution.

To continue the derivation of Eqs.~(\ref{exactGowdyGKE}) we
next define $ r = \sqrt{v - w}$ and
\begin{eqnarray}
a & = & \left\{ \begin{array}{ll}
1    & \mbox{if $\partial_\tau Q = 0$ and
$\partial_\tau P \geq 0$} \\
0    & \mbox{if $\partial_\tau Q = 0$ and
$\partial_\tau P < 0$} \\
|{e^P \partial_\tau Q \over r \, \sqrt{2v}}| &
\mbox{otherwise}
\end{array} \right. , \\
b & = & \left\{ \begin{array}{ll}
0 & \mbox{if $\partial_\tau Q = 0$ and
$\partial_\tau P = 0$} \\
{r \over \sqrt{2v}} & \mbox{otherwise}
\end{array} \right.  .
\end{eqnarray}
Note that $a^2 + b^2  = 1$.  Consider the orthonormal spatial frame
\begin{eqnarray}
\label{orth2a}
V_1 & = & U_1 ,\\
V_2 & = & {e^{(-P+\tau )/ 2} \over \sqrt{\sigma}}
\{(a - b \, e^P \, Q) \partial_x + b \, e^P \partial_y\} ,\\
V_3 & = & {e^{(P+\tau) / 2} \over \sqrt{\sigma}}
\{-( a \, Q + b \, e^{-P} ) \partial_x +a \, \partial_y\}.
\end{eqnarray}
The components of the extrinsic curvature in this frame are
\begin{equation}
k_{ab} = {1 \over 2}{e^{(-\lambda + 3 \tau)/4} }
\left( \begin{array}{ccc}
{(\partial_\tau \lambda- 2 \epsilon+1 )/ 2} &\ \ 
-b \, \sqrt{\epsilon} &\ \  -a \sqrt{\epsilon} \\
-b \, \sqrt{\epsilon} & v-1 & 0 \\
-a \sqrt {\epsilon} & 0 &\  -v-1
\end{array} \right).
\end{equation}
If the twist constant $K$ vanishes (so the spacetime is Gowdy)
then $\epsilon$ vanishes, so the extrinsic curvature is diagonal
in this frame, and Eqs. (\ref{exactGowdyGKE}) follow.  If the
twist constant $K$ does not vanish, the off diagonal components
of the extrinsic curvature are small except during a twist bounce.

The eigenvalues, $\xi_i$, and eigenvectors, $W_i$, of the
extrinsic curvature are the solutions of
${k}_{ab} W_i^b = \xi_i {h}_{ab} W_i^b$.  Perturbation theory
for linear operators \cite{kato} shows that, when $\epsilon$ is
small at some point in space, the difference between the
eigenvalues and the diagonal components and the angles
between the eigenvectors and the frame vectors are both bounded
in terms of $\epsilon$ at that point in space.  This gives a
bound, $50 \sqrt{\epsilon}$, for the magnitude of the
error in Eqs. (\ref{exactGowdyGKE}).  This bound is not
sharp, and holds whether or not the diagonal components of
the extrinsic curvature are well separated from each other.

The eigenvectors of the extrinsic curvature are called the
Kasner directions, or the principal axes.  When $\epsilon$ is
small, the Kasner directions are essentially given by the frame
vectors, $V_i$, in that the angle between each frame vector and
one of the Kasner directions is small.  In the solutions of the
subhamiltonian $H_0 + H_{twist}$, $\epsilon$ grows and decays
again.  We can explicitly compute the rotation of the Kasner
directions with respect to the orthonormal frame, $V_i$ in the
solutions to this subhamiltonian.  Note that $V_1$ is orthogonal
to the $T^2$ isometry orbits and the other two frame vectors are
tangent to the isometry orbits.  We find that in each possible
twist bounce one of the Kasner directions rotates from tangent
to orthogonal, and another rotates from orthogonal to tangent.
Note that each solution to $H_0 + H_{twist}$ is a one parameter
family of Kasner spacetimes.  In this case we can verify directly
that the generalized Kasner exponents reduce to a one parameter
family of Kasner exponents, constant in time.  Since each solution
to the subhamiltonian $H_0 + H_{kin} + H_{twist}$ is also a one
parameter family of Kasners, it must also be the case that during
the evolution governed by this subhamiltonian the generalized
Kasner exponents reduce to a one parameter family of Kasner
exponents, constant in time.  The bounce rule (71) shows that
in this case also, one of the Kasner directions rotates from
being tangent to the isometry orbits to orthogonal, and a
another Kasner direction rotates from orthogonal to tangent.

\section*{Figure Captions}
Fig.~1. Relations between $w$, $w'$, $u$, and $u'$. (a) Curvature
bounce:
Initially, $w > 1$ so that $v = w$. The dashed line shows $v' = |w'| =
|2-w|$. Table II is used to compute $u$ for $v > 1$ and $u'$ for $0 \le
v'
$. The horizontal lines show $1 \le u,u' \le 2$. (b) Twist bounce:
Initially, $-1 \le w \le 3$. The corresponding $u$, $v$, $w'$, $u'$, and
$v'$ are shown. Note that the curves for $u$ and $u'=u$ are superposed.
Table II is used to compute $u$ from $v$.

\medskip

Fig.~2. The local MSS in the $\beta_\pm/|\Omega|$ plane. The triangle
represents the Bianchi IX MSS potential. Scaling the anisotropy
variables
$\beta_\pm$ by the logarithmic volume $|\Omega|$ ($\Omega \to -\infty$
is
the singularity), keeps the bounce locations fixed. The Gowdy models
begin generically with the walls $C$ and $C'$ which disappear as the
spacetime becomes AVTD. In magnetic Gowdy, a third curvature-like wall
$M$ is created by the magnetic field. In $T^2$ symmetric models, a
centrifugal wall $T$ closes off the potential. Here the kinetic ``wall''
is understood to map $C'$ onto $C$ so that the dynamics is confined to
the shaded region. 

\medskip

Fig.~3. Typical behavior of $w(\tau)$ at representative values of
$\theta$. The upper points are offset by 20 and 40 respectively for
display convenience. The flat
regions characterize the Kasner epochs.

\medskip

Fig.~4. Onset and recurrence of the KEA, as monitored by $w$. (a)
$w(\tau,\theta_0)$ for fixed $\theta_0$. Flat regions indicate KEA. (b)
and (c) Detailed closeups of $w(\tau,\theta_0)$ at early (b) and later
(c) times. At the later time, $w$ is flatter during the Kasner epoch.

\medskip

Fig.~5. Typical behavior and accuracy of the bounce laws at three
adjacent spatial points. Each point represents the smallest difference
between a predicted and measured value of $w$ using all of the bounce
rules. This shows that each sequence starts with a twist bounce
and is followed by alternating kinetic and curvature bounces. As the
simulation evolves, the accuracy of the bounce law prediction improves.
The data were obtained by measuring $w$ for all bounces over a symmetric
region (of length $\pi$) in the simulations. The bounces are numbered
consecutively ($N$) following bounces at increasing $\tau$ at a given
point and then moving to the sequence of bounces at the next point. The
vertical lines divide the bounces at a given spatial point from those at
the next point.

\medskip

Fig.~6. Twist bounces identified by the difference between the measured
and predicted values of $w$. All bounces at all spatial points in the
considered interval had their preceding and subsequent values of $w$
measured and then computed according to the twist bounce rule. Where the
difference between the measured and predicted values are large, it means
that the bounce was not a twist bounce. The twist bounces early in the
simulation agree with the bounce rule with rather low accuracy because
the KEA is not yet completely valid. The highest accuracy agreement with
predictions indicates second twist bounces later in the simulation.
Clustering of the more accurate twist bounces just indicates that
similar
behavior is occurring at nearby spatial points.

\medskip

Fig.~7. Behavior of $P$ at twist bounces. Since in a Kasner epoch,
$\partial_\tau P = w$, the piecewise constant $w$ implies a piecewise
linear $P$. The twist bounces are indicated by the arrows. If $w_0 > 1$,
then $w' > 3$ and the next bounce will be a curvature bounce. If $w_0 <
1$, then $w' < -1$ and the next bounce is kinetic.

\medskip

Fig.~8. Combined curvature and twist bounces. This graph is generated as
in Fig.~6 but using the rule for combined bounces. The actual combined
bounces have $|\Delta w| \approx 10^{-6}$.

\medskip

Fig.~9. Structure of a combined bounce. The combined bounce's $w(\tau)$
at $\theta_0$ is shown as a solid line. The segment after the bounce has
$w < -1$ so a kinetic bounce will follow; $w(\tau)$ for a point with
$\theta$ slightly less than $\theta_0$ is shown with the dotted line.
Here the final pre-kinetic bounce segment ($K$) is preceded by pre-twist
($T$) and pre-curvature ($C$) segments. The dot-dashed line shows
$w(\tau)$ for $\theta$ slightly greater than $\theta_0$. Here first a
pre-curvature bounce and then a pre-twist bounce segment precedes the
final pre-kinetic bounce segment.

\medskip

Fig.~10. Spatial resolution dependence. The evolution (for
$w(\theta,\tau)$) is shown
at three representative values of $\theta$ (offset by 15 and 30
respectively) for 1024 (solid line) and 2048 (dotted line with circles)
spatial grid points. The dependence on spatial resolution increases with
the value of
$|w|$ which follows the initial twist bounce.

\medskip

Fig.~11. $|\Delta w|$ for a sequence of bounces at the same value of
$\theta$ for different spatial resolutions. The difference between the
measured and predicted values of $w$ is shown.

\medskip

Fig.~12.  Resolution dependence of waveforms. As has been noted in the
evolution of $U(1)$ symmetric cosmologies \cite{beverly-vinceU(1)}, narrowing
spiky features \cite{numinv} cause the simulations to yield resolution
dependent results where the functions are not smooth. The choice of initial
data made here yields an especially spiky waveform for $P$. A representative
portion is shown. 

\medskip

Fig.~13. Exceptional points. Exceptional points with $\partial_\theta Q
=
0$ and $\partial_\theta P = 0$ are associated with the peaks in $P$
while
$\pi_Q=0$ causes apparent discontinuities in $Q$ \cite{numinv,gowdyphem}. Zero
crossings of all three functions are shown at a late $\tau$ value for a
portion of the $\theta$-axis.

\medskip

Fig.~14. The number of exceptional points vs $\tau$. The growth in the
number of exceptional points vs $\tau$ is shown. While $N$ appears to
level off, this could just reflect the exponential increase of Kasner
epoch duration characteristic of mixmaster dynamics. $N(\tau)$ is {\it
not} a power law.

\medskip

Fig.~15. $P(\theta,\tau)$ (top), $Q(\theta,\tau)$ (middle), and
$\lambda(\theta,\tau)$ (bottom) are shown for the full simulation (with
arbitrary scales for their values). The left hand column uses 1024 and
the right 2048 spatial grid points. In each frame, the horizontal axis
is
$-\pi/5 \le \theta \le 9\pi/5$ and the vertical axis $0 \le \tau \le 76$.

\medskip

Fig.~16. Limits on the simulation. The plot shows $w'(\theta,\tau)$
computed from the simulations $w(\theta,\tau)$ using the twist bounce
rule. The scale is set so that values $> 100$ ( $< -100$) appear white
(black). The white and black lines which extend to the end of the
simulation indicate $\theta$-values which are destined to have
dangerously large values of $w$ after the next twist bounce. Only a
portion of the $\theta$ axis is shown.

\newpage

\begin{table}
\caption{Behavior of Potentials when KEA
holds.}
\begin{center}
\renewcommand{\arraystretch}{2}
\begin{tabular}{|ccc|lll|llll|} \hline
\multicolumn{3}{|c}{Condition} \vline & 
\multicolumn{3}{c}{Potentials that Grow} \vline &
\multicolumn{4}{c}{Potentials that do not Grow} \vline \\ \hline
$\; \; \; w \leq -1$ & and & $\pi_Q \neq 0$ &  $H_{kin}$ & & & 
& $H_{small}$ &  $H_{curv}$ & $H_{twist}$ \\ \hline
$-1 \leq w < 0$ & and & $\pi_Q \neq 0$ & $H_{kin}$& & $H_{twist}$ &
& $H_{small}$ & $H_{curv}$ & \\ \hline
$0 \leq w \leq 1$ & &  & & & $H_{twist}$ &
$H_{kin}$ & $H_{small}$ & $H_{curv}$ & \\ \hline
$1 < w < 3$ & and & $\partial_\theta Q \neq 0$  & & $H_{curv}$ &
$H_{twist}$ &
$H_{kin}$ & $H_{small}$ & & \\ \hline
$3 \leq w \; \; \;$ & and & $\partial_\theta Q \neq 0$ & & $H_{curv}$ &
&
$H_{kin}$ & $H_{small}$ & & $H_{twist}$ \\ \hline
\end{tabular}
\end{center}
\label{tpotential}
\end{table}
\begin{table}
\renewcommand{\arraystretch}{2}
\caption{Calculation of BKL parameter $u$.}
\begin{center}
\begin{tabular}{|l||c|c|c|} \hline
Range of $v$ & $0 \leq v < 1$ & $1 < v \leq 3$
& $3 \leq v$ \\ \hline
Value of $u$ & $u = {1+v \over 1-v}$
& $u = {2 \over v-1}$
& $u = {v-1 \over 2}$ \\ \hline
\end{tabular}
\end{center}
\label{uval}
\end{table}
\begin{table}
\caption{Summary of bounce rules.}
\begin{tabular}{|l||c|c|c|c|c|}
$\ $Bounce type$\ $&$\
$Kinetic&Curvature$\ $&$\ $Twist&$\ $Curvature-Twist$\ $&$\ $Kinetic-Twist$\
$\\
\tableline
$\ $&$\ $&$\ $&$\ $&$\ $&\\
Bounce rule&$w'=-w\ \ $&$w' = 2-w\ \ $&$w' = {{w+3} \over {w-1}}\ \ $&$w' =
{{w-5}
\over {w-1}}\ \ $&$w' = {{3-w} \over {w+1}}\ \ $ 
\end{tabular}
\label{bounceruletable}
\end{table}

\begin{figure}[bth]
\begin{center}
\makebox[4in]{\psfig{file=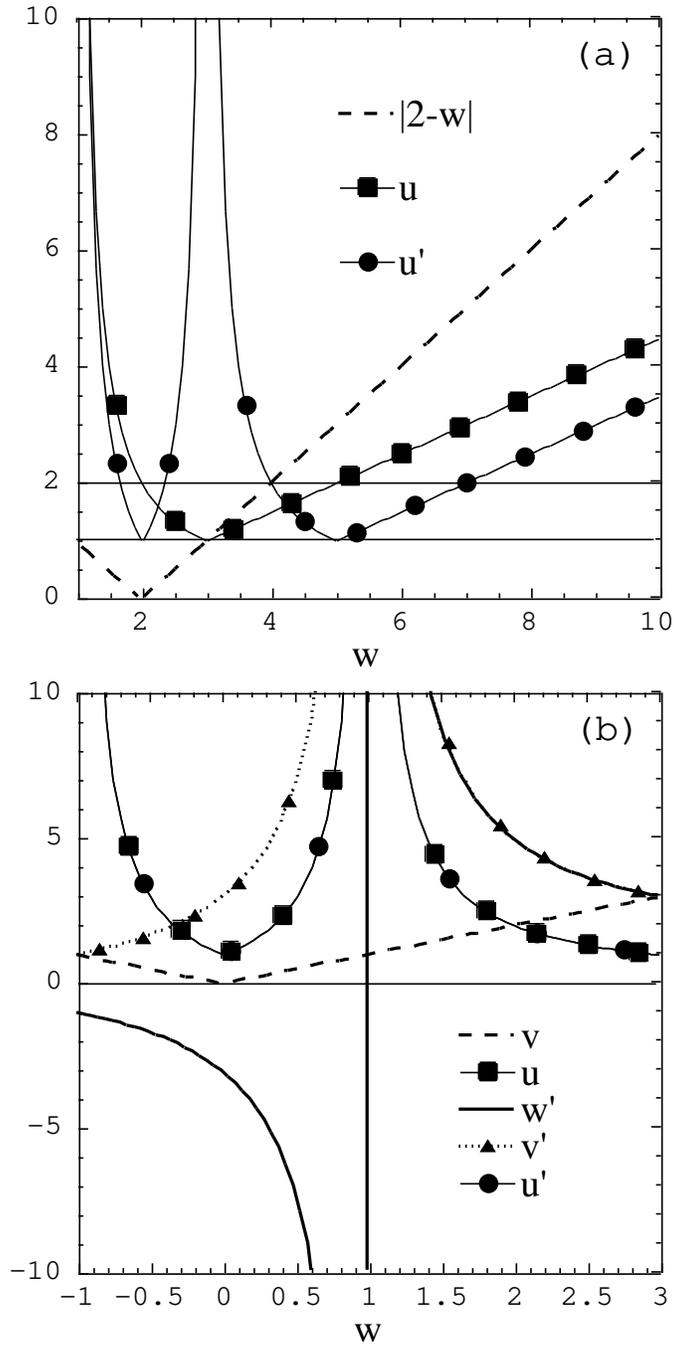,width=3.5in}}
\caption{Relations between $w$, $w'$, $u$, and $u'$. (a) Curvature
bounce:
Initially, $w > 1$ so that $v = w$. The dashed line shows $v' = |w'| =
|2-w|$. Table II is used to compute $u$ for $v > 1$ and $u'$ for $0 \le
v'
$. The horizontal lines show $1 \le u,u' \le 2$. (b) Twist bounce:
Initially, $-1 \le w \le 3$. The corresponding $u$, $v$, $w'$, $u'$, and
$v'$ are shown. Note that the curves for $u$ and $u'=u$ are superposed.
Table II is used to compute $u$ from $v$.
}
\end{center}
\end{figure}
\begin{figure}[bth]
\begin{center}
\makebox[4in]{\psfig{file=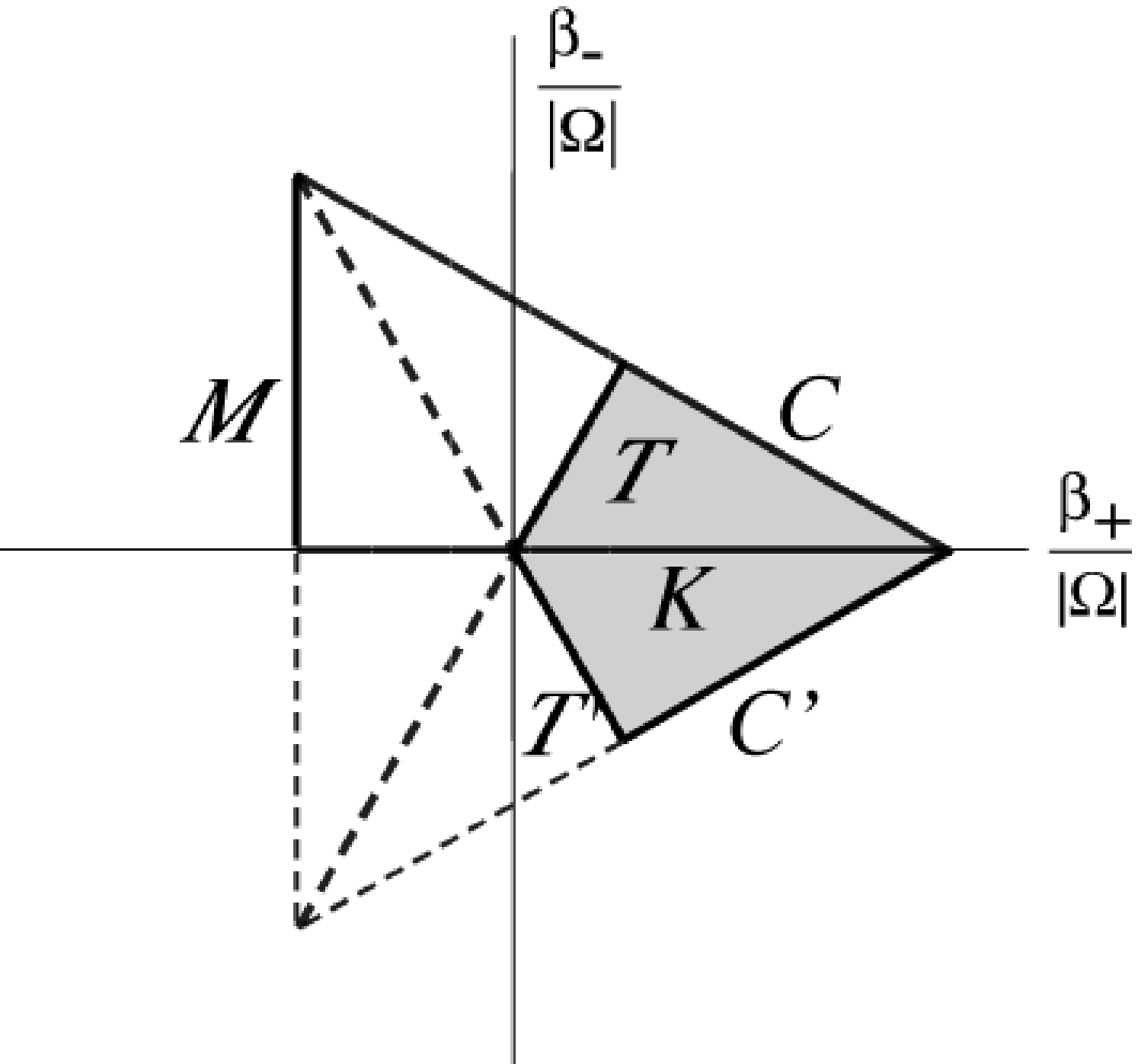,width=3.0in}}
\caption{Fig.~2. The local MSS in the $\beta_\pm/|\Omega|$ plane. The triangle
represents the Bianchi IX MSS potential. Scaling the anisotropy
variables
$\beta_\pm$ by the logarithmic volume $|\Omega|$ ($\Omega \to -\infty$
is
the singularity), keeps the bounce locations fixed. The Gowdy models
begin generically with the walls $C$ and $C'$ which disappear as the
spacetime becomes AVTD. In magnetic Gowdy, a third curvature-like wall
$M$ is created by the magnetic field. In $T^2$ symmetric models, a
centrifugal wall $T$ closes off the potential. Here the kinetic ``wall''
is understood to map $C'$ onto $C$ so that the dynamics is confined to
the shaded region. }
\end{center}
\end{figure}
\begin{figure}[bth]
\begin{center}
\makebox[4in]{\psfig{file=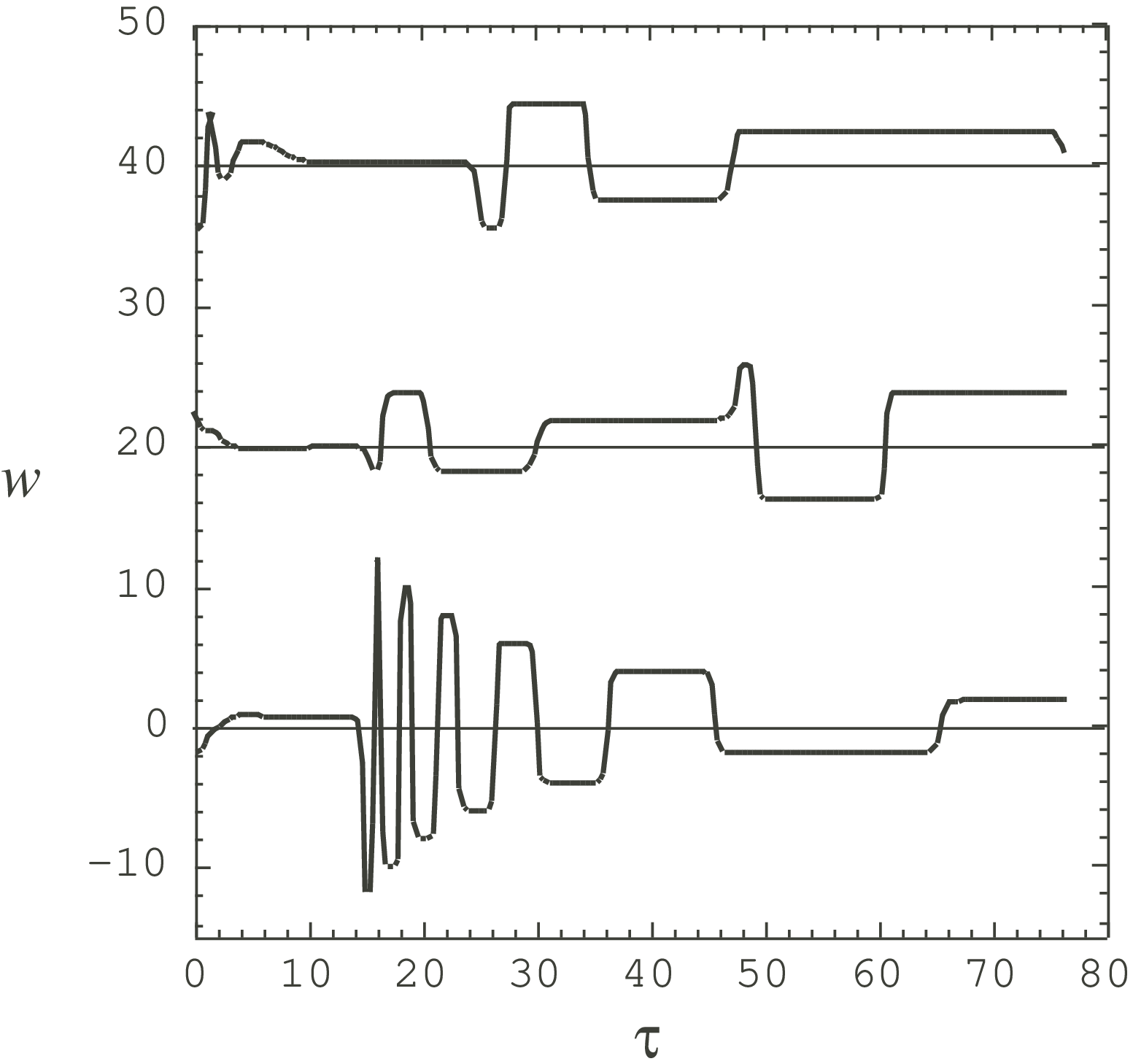,width=3.5in}}
\caption{Fig.~3. Typical behavior of $w(\tau)$ at representative values of
$\theta$. The upper points are offset by 20 and 40 respectively for
display convenience. The flat
regions characterize the Kasner epochs.
}
\end{center}
\end{figure}
\begin{figure}[bth]
\begin{center}
\makebox[4in]{\psfig{file=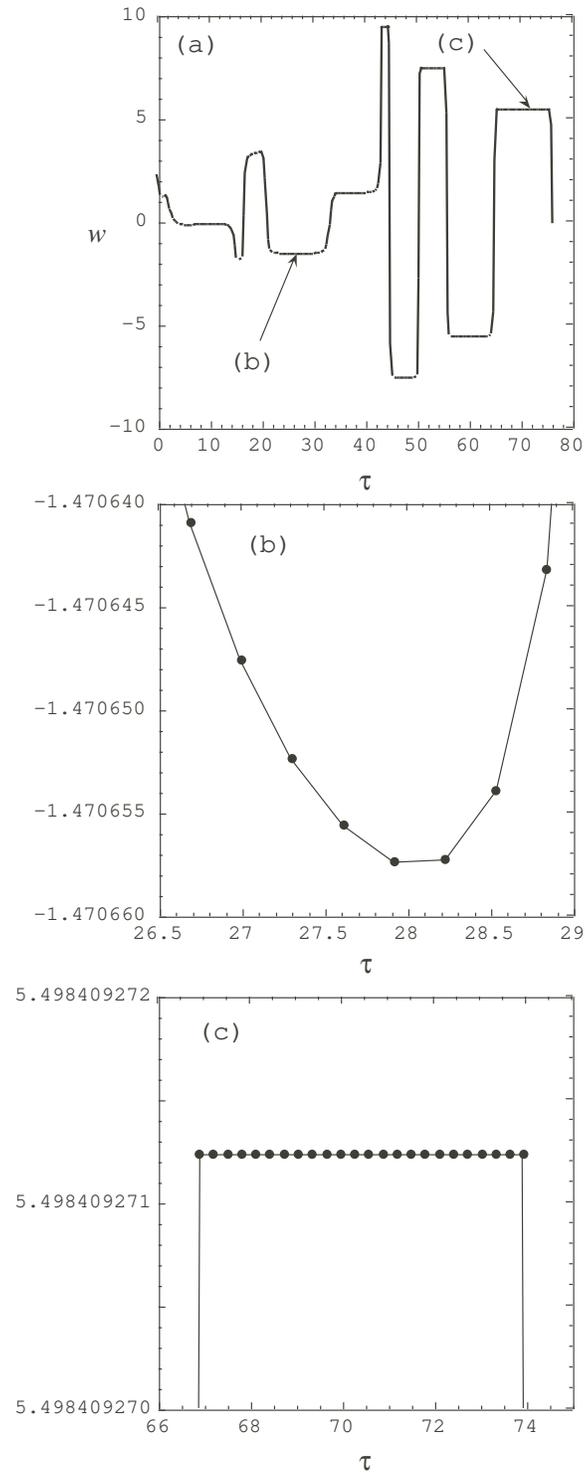,width=3.0in}}
\caption{Onset and recurrence of the KEA, as monitored by $w$. (a)
$w(\tau,\theta_0)$ for fixed $\theta_0$. Flat regions indicate KEA. (b)
and (c) Detailed closeups of $w(\tau,\theta_0)$ at early (b) and later
(c) times. At the later time, $w$ is flatter during the Kasner epoch.}
\end{center}
\end{figure}
\begin{figure}[bth]
\begin{center}
\makebox[4in]{\psfig{file=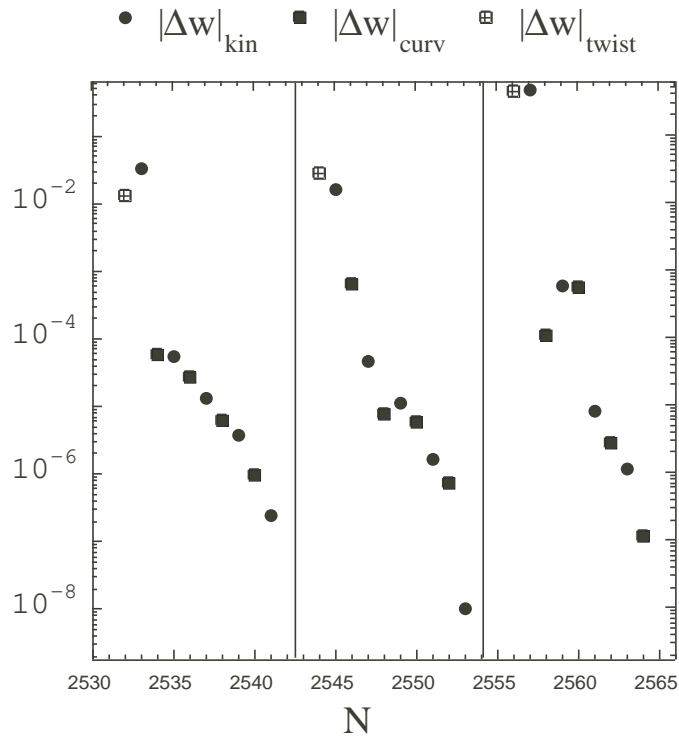,width=3.5in}}
\caption{Typical behavior and accuracy of the bounce laws at three
adjacent spatial points. Each point represents the smallest difference
between a predicted and measured value of $w$ using all of the bounce
rules. This shows that each sequence starts with a twist bounce
and is followed by alternating kinetic and curvature bounces. As the
simulation evolves, the accuracy of the bounce law prediction improves.
The data were obtained by measuring $w$ for all bounces over a symmetric
region (of length $\pi$) in the simulations. The bounces are numbered
consecutively ($N$) following bounces at increasing $\tau$ at a given
point and then moving to the sequence of bounces at the next point. The
vertical lines divide the bounces at a given spatial point from those at
the next point.
}
\end{center}
\end{figure}
\begin{figure}[bth]
\begin{center}
\makebox[4in]{\psfig{file=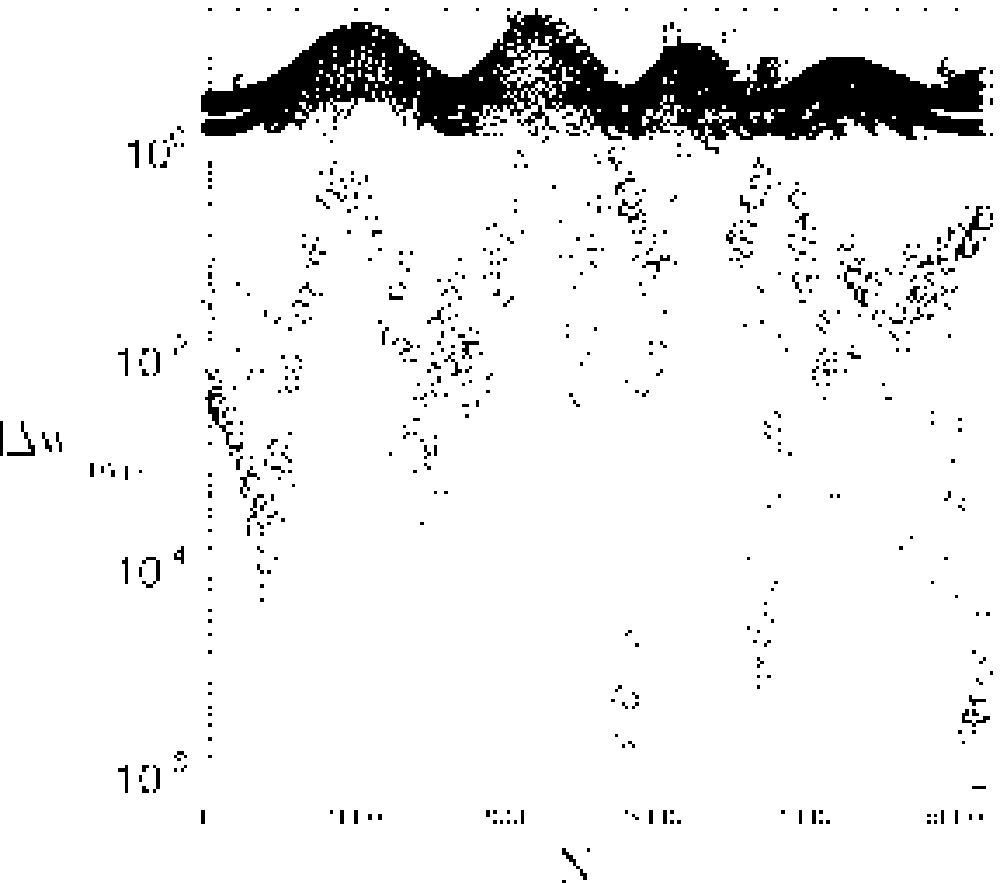,width=3.5in}}
\caption{Twist bounces identified by the difference between the measured
and predicted values of $w$. All bounces at all spatial points in the
considered interval had their preceding and subsequent values of $w$
measured and then computed according to the twist bounce rule. Where the
difference between the measured and predicted values are large, it means
that the bounce was not a twist bounce. The twist bounces early in the
simulation agree with the bounce rule with rather low accuracy because
the KEA is not yet completely valid. The highest accuracy agreement with
predictions indicates second twist bounces later in the simulation.
Clustering of the more accurate twist bounces just indicates that
similar
behavior is occurring at nearby spatial points.
}
\end{center}
\end{figure}
\begin{figure}[bth]
\begin{center}
\makebox[4in]{\psfig{file=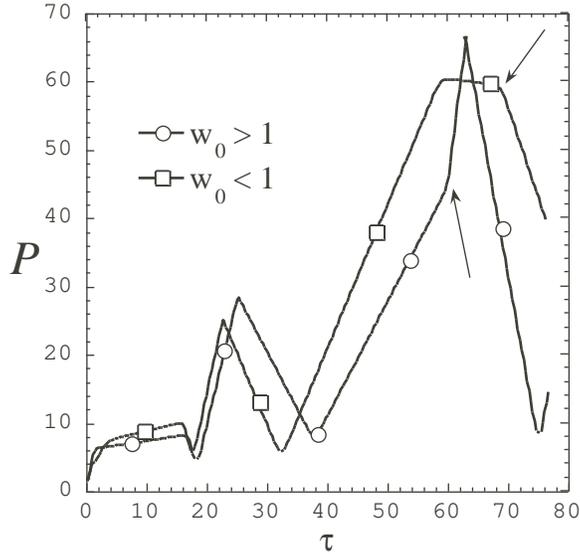,width=3.0in}}
\caption{Behavior of $P$ at twist bounces. Since in a Kasner epoch,
$\partial_\tau P = w$, the piecewise constant $w$ implies a piecewise
linear $P$. The twist bounces are indicated by the arrows. If $w_0 > 1$,
then $w' > 3$ and the next bounce will be a curvature bounce. If $w_0 <
1$, then $w' < -1$ and the next bounce is kinetic.
}
\end{center}
\end{figure}
\begin{figure}[bth]
\begin{center}
\makebox[4in]{\psfig{file=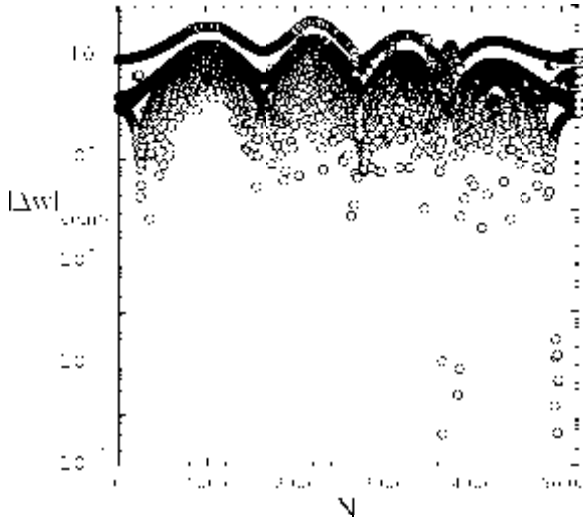,width=3.0in}}
\caption{Combined curvature and twist bounces. This graph is generated as
in Fig.~6 but using the rule for combined bounces. The actual combined
bounces have $|\Delta w| \approx 10^{-6}$.
}
\end{center}
\end{figure}
\begin{figure}[bth]
\begin{center}
\makebox[4in]{\psfig{file=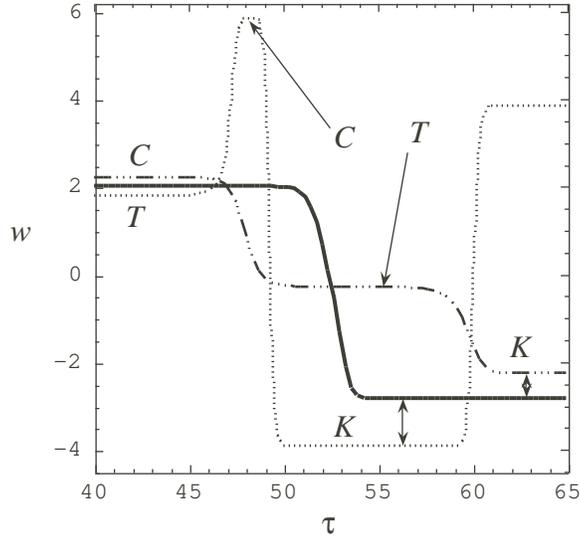,width=3.0in}}
\caption{Structure of a combined bounce. The combined bounce's $w(\tau)$
at $\theta_0$ is shown as a solid line. The segment after the bounce has
$w < -1$ so a kinetic bounce will follow; $w(\tau)$ for a point with
$\theta$ slightly less than $\theta_0$ is shown with the dotted line.
Here the final pre-kinetic bounce segment ($K$) is preceded by pre-twist
($T$) and pre-curvature ($C$) segments. The dot-dashed line shows
$w(\tau)$ for $\theta$ slightly greater than $\theta_0$. Here first a
pre-curvature bounce and then a pre-twist bounce segment precedes the
final pre-kinetic bounce segment.
}
\end{center}
\end{figure}
\begin{figure}[bth]
\begin{center}
\makebox[4in]{\psfig{file=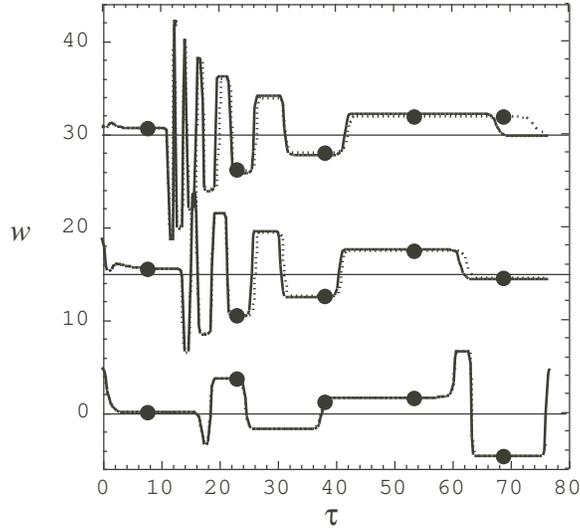,width=3.0in}}
\caption{Spatial resolution dependence. The evolution (for
$w(\theta,\tau)$) is shown
at three representative values of $\theta$ (offset by 15 and 30
respectively) for 1024 (solid line) and 2048 (dotted line with circles)
spatial grid points. The dependence on spatial resolution increases with
the value of
$|w|$ which follows the initial twist bounce.
}
\end{center}
\end{figure}
\begin{figure}[bth]
\begin{center}
\makebox[4in]{\psfig{file=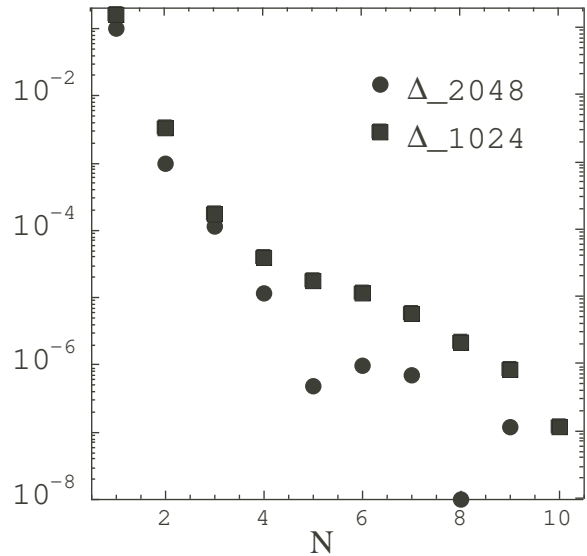,width=3.0in}}
\caption{$|\Delta w|$ for a sequence of bounces at the same value of
$\theta$ for different spatial resolutions. The difference between the
measured and predicted values of $w$ is shown.
}
\end{center}
\end{figure}
\begin{figure}[bth]
\begin{center}
\makebox[4in]{\psfig{file=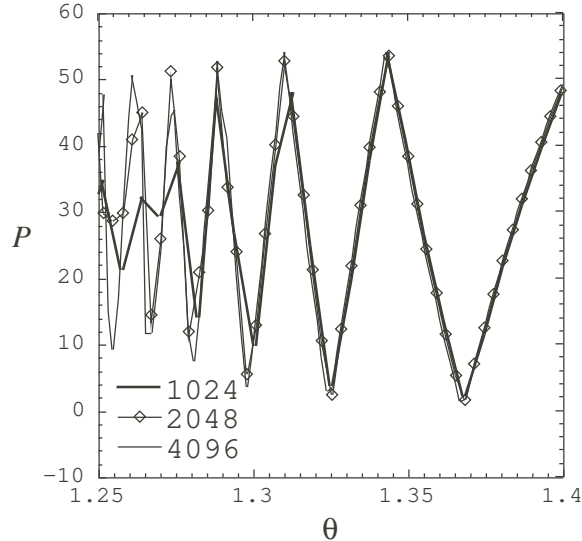,width=3.0in}}
\caption{Resolution dependence of waveforms. As has been noted in the
evolution of $U(1)$ symmetric cosmologies \protect \cite{beverly-vinceU(1)},
narrowing spiky features \protect \cite{numinv} cause the simulations to yield
resolution dependent results where the functions are not smooth. The choice
of initial data made here yields an especially spiky waveform for $P$. A
representative portion is shown. 
}
\end{center}
\end{figure}
\begin{figure}[bth]
\begin{center}
\makebox[4in]{\psfig{file=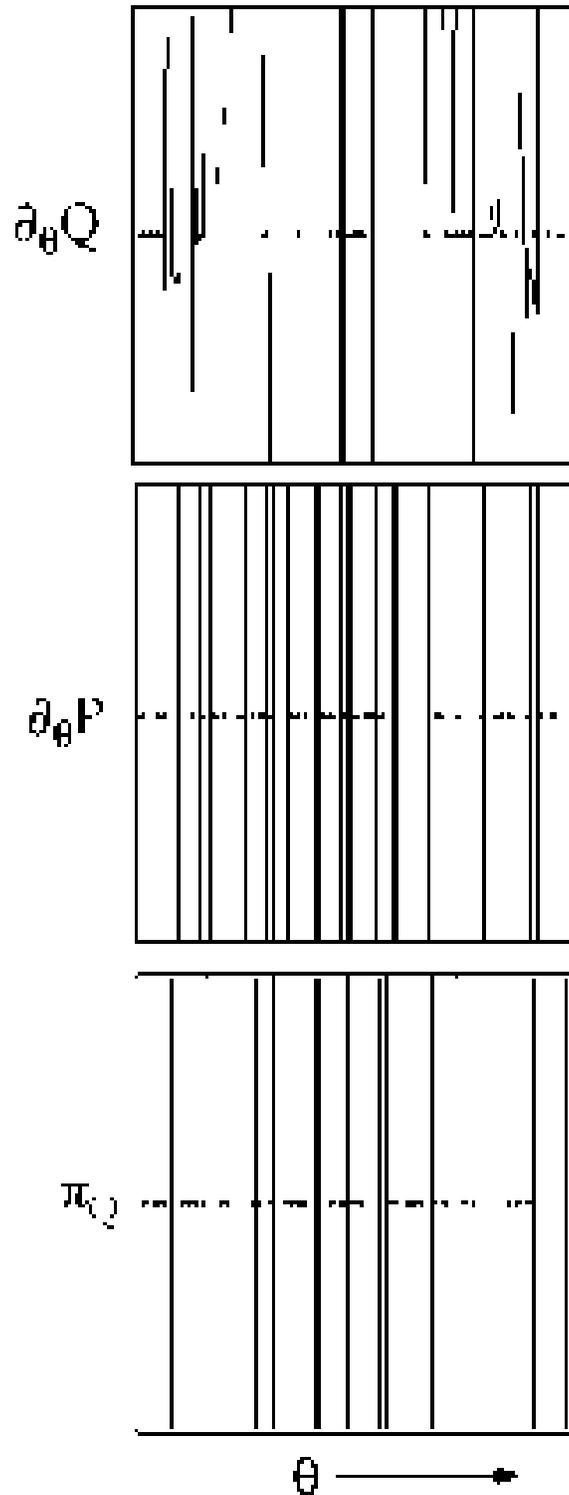,width=3.0in}}
\caption{Exceptional points. Exceptional points with $\partial_\theta Q
=
0$ and $\partial_\theta P = 0$ are associated with the peaks in $P$
while
$\pi_Q=0$ causes apparent discontinuities in $Q$
\protect \cite{numinv,gowdyphem}. Zero crossings of all three functions are
shown at a late $\tau$ value for a portion of the $\theta$-axis.
}
\end{center}
\end{figure}
\begin{figure}[bth]
\begin{center}
\makebox[4in]{\psfig{file=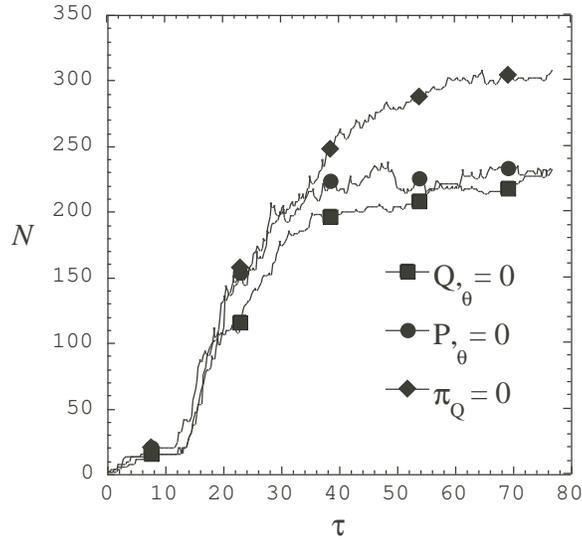,width=3.0in}}
\caption{The number of exceptional points vs $\tau$. The growth in the
number of exceptional points vs $\tau$ is shown. While $N$ appears to
level off, this could just reflect the exponential increase of Kasner
epoch duration characteristic of mixmaster dynamics. $N(\tau)$ is {\protect
\it not} a power law.
}
\end{center}
\end{figure}
\begin{figure}[bth]
\begin{center}
\makebox[4in]{\psfig{file=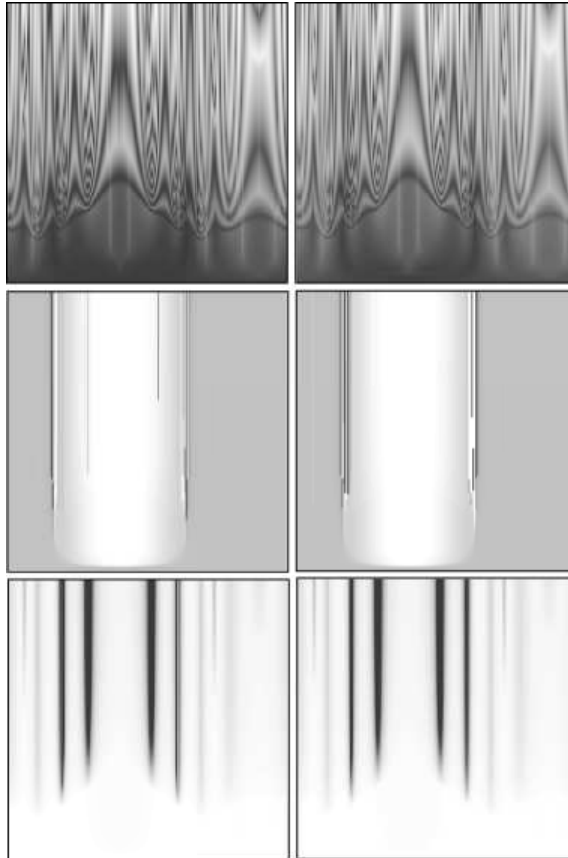,width=3.0in}}
\caption{$P(\theta,\tau)$ (top), $Q(\theta,\tau)$ (middle), and
$\lambda(\theta,\tau)$ (bottom) are shown for the full simulation (with
arbitrary scales for their values). The left hand column uses 1024 and
the right 2048 spatial grid points. In each frame, the horizontal axis
is
$-\pi/5 \le \theta \le 9\pi/5$ and the vertical axis $0 \le \tau \le 76$.}
\end{center}
\end{figure}
\begin{figure}[bth]
\begin{center}
\makebox[4in]{\psfig{file=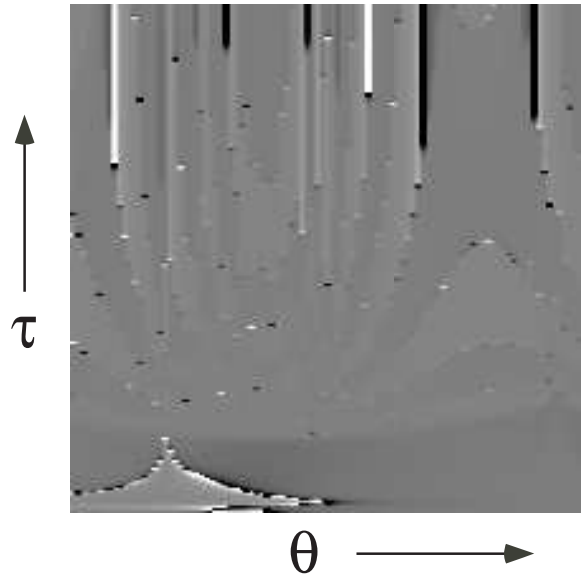,width=3.0in}}
\caption{Limits on the simulation. The plot shows $w'(\theta,\tau)$
computed from the simulations $w(\theta,\tau)$ using the twist bounce
rule. The scale is set so that values $> 100$ ( $< -100$) appear white
(black). The white and black lines which extend to the end of the
simulation indicate $\theta$-values which are destined to have
dangerously large values of $w$ after the next twist bounce. Only a
portion of the $\theta$ axis is shown.
}
\end{center}
\end{figure}

\end{document}